\documentclass[12pt,draftcls,onecolumn]{IEEEtran}

\usepackage[dvips]{color}
\usepackage{graphicx}
\usepackage{amsmath}
\usepackage{color}

\begin{document}

\title{Secrecy Outage Analysis of Non-Orthogonal Spectrum Sharing for Heterogeneous Cellular Networks}

\markboth{IEEE Transactions on Communications (under review)}%
{Yulong~Zou \emph{et al.}: Secrecy Outage Analysis of Non-Orthogonal Spectrum Sharing for Heterogeneous Cellular Networks}

\author{Yulong~Zou, Tong Wu, Ming~Sun, Jia~Zhu, Mujun~Qian, and Chen Liu


\thanks{The authors are with the Nanjing University of Posts and Telecommunications, Nanjing, Jiangsu, China.}



}

\maketitle

\vspace{-1cm}

\begin{abstract}
In this paper, we investigate physical-layer security for a heterogeneous cellular network consisting of a macro cell and a small cell in the presence of a passive eavesdropper that intends to tap both the macro-cell and small-cell transmissions. Both the orthogonal spectrum sharing (OSS) and non-orthogonal spectrum sharing (NOSS) are considered for the heterogeneous cellular network. The OSS allows the macro cell and small cell to access a given spectrum band in an orthogonal manner, whereas the NOSS enables them to access the same spectrum simultaneously and mutual interference exits. As a consequence, we present two NOSS schemes, namely the interference-limited NOSS (IL-NOSS) and interference-canceled NOSS (IC-NOSS), where the mutual interference is constrained below a tolerable level in the IL-NOSS through power control and the IC-NOSS scheme exploits a specially-designed signal for canceling out the interference received a legitimate cellular user while confusing the passive eavesdropper. We derive closed-form expressions for an overall secrecy outage probability of the OSS, IL-NOSS, and IC-NOSS schemes, which take into account the transmission secrecy of both the macro cell and small cell. We further characterize the secrecy diversity of OSS, IL-NOSS and IC-NOSS schemes through an asymptotic secrecy outage analysis in the high signal-to-noise ratio region. It is shown that the OSS and IL-NOSS methods obtain the same secrecy diversity gain of zero only, and a higher secrecy diversity gain of one is achieved by the IC-NOSS scheme. Additionally, numerical results demonstrate that the IC-NOSS scheme significantly performs better than the OSS and IL-NOSS methods in terms of the overall secrecy outage probability.
\end{abstract}


\section{Introduction}
\IEEEPARstart{D}{riven} by an explosive increase of wireless data traffic, dense small base stations with low power are deployed in a macro cell to increase the network capacity and accommodate more user terminals [1]-[4]. Such a network architecture is termed as the heterogeneous cellular network, where a macro cell typically coexists with various small cells, e.g., femto cells, pico cells and micro cells [5]. In heterogeneous cellular networks, a number of economical small base stations (SBSs) are generally deployed around a relatively expensive macro base station (MBS), which can expand the network capacity and coverage at a reduced expense [6], [7]. Meanwhile, there are also some challenging issues to be addressed for heterogeneous cellular networks, including the user association, resource allocation and interference management [8]-[10]. For example, the authors of [11] studied energy efficiency and load balancing for the user association problem, and proposed an energy conservation algorithm for energy-efficient transmissions. In [12], a load-balancing scheme was presented for addressing the association between subscribers and available wireless local area networks (WLANs).

Typically, both the orthogonal and non-orthogonal spectrum sharing may be considered for heterogeneous cellular networks, where a number of small cells are in the range of macro cells [13]. In the orthogonal spectrum sharing mechanism, a given spectrum band of heterogeneous networks is first divided into a series of orthogonal sub-bands, which are then allocated to the macro cells and small cells, respectively. In other words, the macro cells and small cells are allowed to exclusively occupy their respective sub-bands, which is referred to as orthogonal spectrum sharing (OSS). By contrast, the non-orthogonal spectrum sharing (NOSS) enables SBSs and MBSs to simultaneously access the same spectrum, also called non-orthogonal multiple access (NOMA) in [14], [15]. As discussed in existing literature [14]-[18], the NOSS can achieve a higher spectral efficiency than OSS, since the same spectrum resources are simultaneously accessed by both the macro cell and small cell in heterogeneous cellular networks. However, this also results in mutual interference between the macro cell and small cell, which may degrade the quality-of-service (QoS) of heterogeneous networks.

To this end, an interference management scheme was presented in [19] for the sake of alleviating the adverse impact of mutual interference on intended users. Moreover, macroscopic and microscopic control schemes were proposed in [20] for the purpose of mitigating intercell interference in heterogeneous networks. It is noted that the broadcast nature of radio propagation and open architecture of heterogeneous networks make legitimate wireless transmissions of both macro cells and small cells become extremely vulnerable to eavesdropping attacks [21]. Therefore, it is particularly important and necessary to explore the transmission security for heterogeneous cellular networks. Differing from traditional encryption technology, which only provides the computational security, physical-layer security (PLS) is emerging as a promising means of guaranteeing perfect secrecy by exploiting physical characteristics of the communications medium [22], {{[23]}}. In the past decades, extensive research efforts have been devoted to improving PLS with the aid of cooperative relaying [24]{{-[27]}}, multiple-input multiple-output (MIMO) [28]{{-[30]}}, beamforming [31], [32], {{and millimeter-wave techniques [33], [34]}}.

Until now, an increasing research attention has been paid to PLS for NOSS systems in heterogeneous cellular networks [35]-[39]. {{To be specific, the PLS for large-scale non-orthogonal multiple access networks was studied in [35] by using stochastic geometry and closed-form expressions for the secrecy outage probability were derived for both single-antenna and multiple-antenna scenarios. In [36], an effect of beamforming and artificial noise generation on the PLS of large-scale spectrum sharing networks was examined in terms of an average secrecy rate and secrecy outage probability. Later on, an access threshold based secrecy mobile association scheme was proposed in [37] for improving the secrecy throughput of heterogeneous networks.}} Moreover, the author of [38] investigated the PLS for orthogonal spectrum sharing systems with the aid of joint multiuser scheduling and friendly jammer selection. It was found in [38] that no extra secrecy diversity gain is achievable through the use of an opportunistic friendly jammer selection. {{In addition, the PLS benefit of heterogeneous ultra-dense networks was studied in [39] by jointly considering the caching and wireless energy harvesting.}} More recently, an interference-canceled underlay spectrum-sharing (IC-USS) scheme, called IC-NOSS throughout this paper, was proposed in [1] for the sake of enhancing the PLS of heterogeneous cellular networks, where closed-form expressions for outage probability and intercept probability are derived.

{{This paper differs from [1] in the following aspects. First, we examine the secrecy outage probability of IC-NOSS scheme in this paper, which is different from [1], where closed-form expressions for intercept probability and outage probability are derived from a security-reliability tradeoff (SRT) perspective. Mathematically speaking, it is more challenging to obtain a closed-form expression for secrecy outage probability than that for the intercept probability and outage probability. Second, this paper characterizes the secrecy diversity of IC-NOSS through an asymptotic secrecy outage analysis in the high signal-to-noise ratio (SNR) region, however no secrecy diversity is provided in [1].}} The main contributions of this paper are summarized as follows. First, we derive closed-form expressions for an overall secrecy outage probability of the IC-NOSS, interference-limited NOSS (IL-NOSS) and OSS schemes by jointly taking into account both the macro-cell and small-cell transmissions. Second, the secrecy diversity analysis of these three schemes is carried out by characterizing an asymptotic behavior of the secrecy outage probability in the high SNR region. Finally, numerical results show that the overall secrecy outage probability of IC-NOSS is substantially smaller than that of the OSS and IL-NOSS methods.

The reminder of this paper is organized as follows. In Section II, we present the system model of spectrum-sharing heterogeneous cellular networks and introduce the OSS, IL-NOSS and IC-NOSS schemes. In Section III, closed-form expressions for the secrecy outage probability of these schemes are derived over Rayleigh fading channels, followed by Section IV, where the secrecy diversity analysis is conducted. Next, numerical results and discussions are provided in Section V. Finally, Section VI gives some concluding remarks.

\section{System Model}
\begin{figure*}
  \centering
  {\includegraphics[scale=0.35]{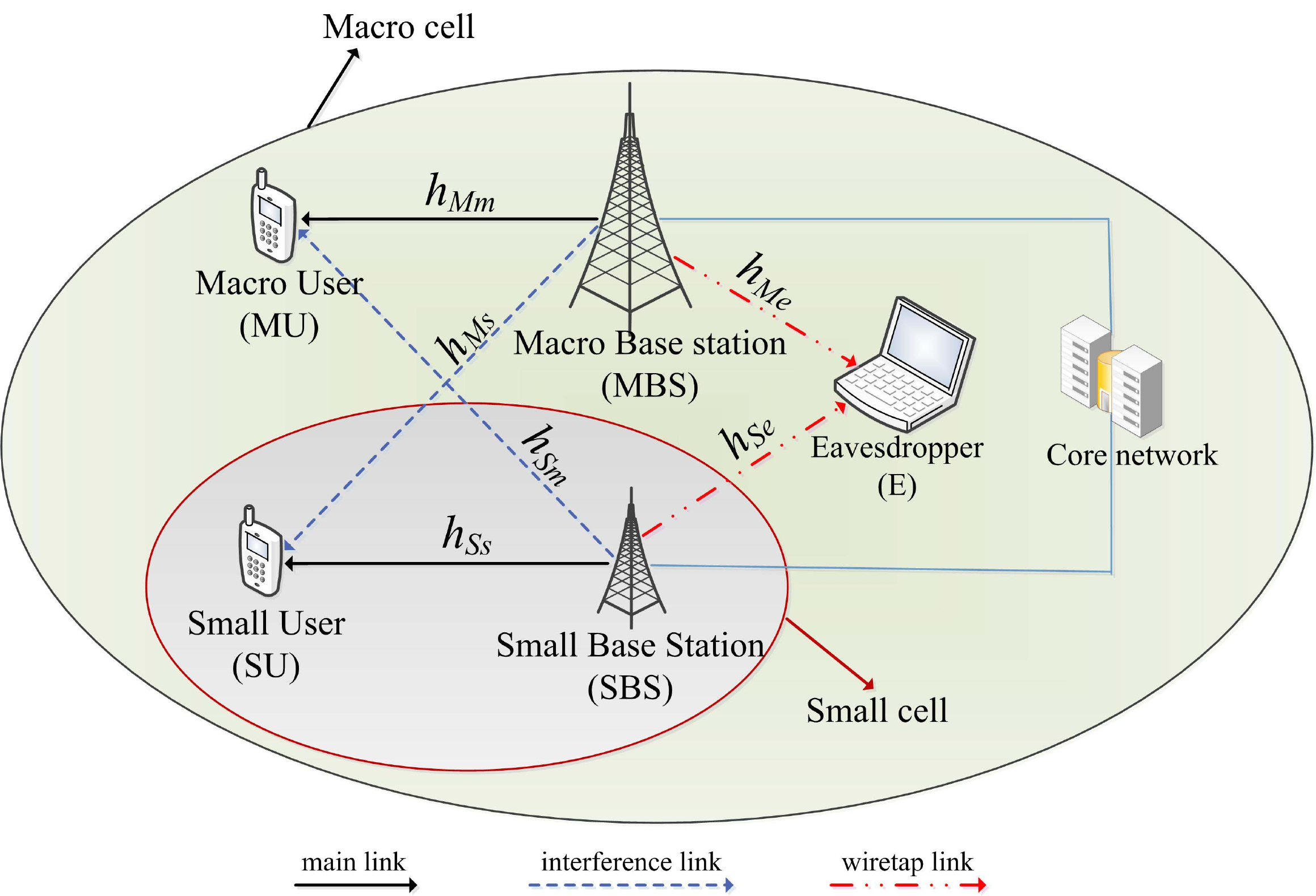}\\
  \caption{A heterogeneous cellular network composed of a macro cell and a small cell in the presence of a common eavesdropper (E).}\label{Fig1}}
\end{figure*}

In this section, we first describe the system model of a heterogeneous cellular network, where a small cell coexists with a macro cell in the face of a common {{passive eavesdropper (E) which is considered to silently wiretap the confidential messages of both the macro cell and small cell}}. As shown in Fig. 1, in the macro cell, MBS transmits its message to a macro user (MU), while SBS communicates with a small user (SU) in the small cell. An eavesdropper is assumed to tap both the MBS-MU and SBS-SU transmissions. Notice that both the MBS and SBS are connected to a core network via fiber cables, e.g., a mobile switch center (MSC) in the global system for mobile communication (GSM), a mobility management entity (MME) in the long term evolution (LTE), which guarantees the real-time information exchange between MBS and SBS. {{Notice that the focus of this paper is to analyze the secrecy outage probability and secrecy diversity of a specific heterogeneous network with a single macro cell and small cell. It is not mathematically tractable to derive closed-form secrecy outage probability for some more complex heterogeneous networks e.g. with randomly distributed multi-antenna nodes. Our derived secrecy outage probability expressions can be used to facilitate the secrecy performance evaluation of such complex networks through Monte-Carlo simulations. Moreover, the aforementioned system model of Fig. 1 is also applicable to a general heterogeneous network consisting of multiple MBSs and SBSs with the aid of base station pairing and grouping. If more than one MBS-SBS pair is available, we may divide the given spectrum into multiple orthogonal sub-bands to be assigned to different MBS-SBS pairs.}}

{{The macro cell and small cell are allowed to share the same spectrum band, and both the OSS and NOSS mechanisms are considered for the heterogeneous cellular network of Fig. 1. In the OSS scheme, the spectrum band is first divided into orthogonal sub-bands, which are then allocated to the macro cell and small cell, respectively. In this way, no mutual interference occurs between the macro cell and small cell. By contrast, in the NOSS mechanism, MBS and SBS simultaneously transmit their respective confidential messages over the same spectrum band. This can enhance the spectral efficiency, which, however, comes at the cost of degrading the QoS of transmissions, since mutual interference between the macro cell and small cell happens in this case.}}

{{In order to guarantee the QoS of heterogeneous cellular networks, transmit powers of MBS and SBS, denoted by $P_M$ and $P_S$, respectively, are constrained in the conventional IL-NOSS method for the sake of keeping the mutual interference below a tolerable level. Differing from the IL-NOSS, the IC-NOSS scheme exploits a specially-designed signal transmitted at MBS, which not only generates a certain interference to the eavesdropper, but can also cancel out the mutual interference received at MU from SBS. Following the existing literature [24]-[32], the channel coefficients of MBS-MU, MBS-SU, SBS-SU, SBS-MU, MBS-E, and SBS-E are, respectively, represented by $h_{Mm}$, $h_{Ms}$, $h_{Ss}$, $h_{Sm}$, $h_{Me}$, and $h_{Se}$, which are modeled as independent Rayleigh fading processes with respective channel gains of $\sigma^2_{Mm}$, $\sigma^2_{Ms}$, $\sigma^2_{Ss}$, $\sigma^2_{Sm}$, $\sigma^2_{Me}$, and $\sigma^2_{Se}$, namely $\sigma^2_{Mm}=E(|h_{Mm}|^2)$, $\sigma^2_{Ms}=E(|h_{Ms}|^2)$, $\sigma^2_{Ss}=E(|h_{Ss}|^2)$, $\sigma^2_{Sm}=E(|h_{Sm}|^2)$, $\sigma^2_{Me}=E(|h_{Me}|^2)$, and $\sigma^2_{Se}=E(|h_{Se}|^2)$. Although only the Rayleigh fading model is considered in this paper, similar secrecy results can be obtained for other channel models, e.g., Nakagami fading [40], Rician fading [41], and $\alpha$-$\mu$ fading [42].}}

{{According to [3] and [43], the aforementioned channel gains are modeled as $\sigma^2_{Mm}= {d^{-\alpha_{Mm}}_{Mm}}\delta _{Mm}^2$, $ \sigma^2_{Ms} = {d^{-\alpha_{Ms}}_{Ms}}\delta _{Ms}^2$, $ \sigma^2_{Ss}= {d^{-\alpha_{Ss}}_{Ss}}\delta _{Ss}^2$, $ \sigma^2_{Sm} = {d^{-\alpha_{Sm}}_{Sm}}\delta _{Sm}^2$, $ \sigma^2_{Me} = {d^{-\alpha_{Me}}_{Me}}\delta _{Me}^2$, and $ \sigma^2_{Se} = {d^{-\alpha_{Se}}_{Se}}\delta _{Se}^2$ to show an effect of the large-scale path loss, where $d_{Mm}$, $d_{Ms}$, $d_{Ss}$, $d_{Sm}$, $d_{Me}$, $d_{Se}$ are transmission distances of the MBS-MU, MBS-SU, SBS-SU, SBS-MU, MBS-E, SBS-E  channels, respectively, and, moreover, $\alpha_{Mm}$, $\alpha_{Ms}$, $\alpha_{Ss}$, $\alpha_{Sm}$, $\alpha_{Me}$, and $\alpha_{Se}$ are their respective path loss factors. Besides, $\delta^2_{Mm}$, $\delta^2_{Ms}$, $\delta^2_{Ss}$, $\delta^2_{Sm}$, $\delta^2_{Me}$, and $\delta^2_{Se}$ represent small-scale fading variances of the MBS-MU, MBS-SU, SBS-SU, SBS-MU, MBS-E and SBS-E channels, respectively.}} Additionally, the additive white Gaussian noise (AWGN) with a zero mean and a variance of $N_0$ is encountered at any receiver of Fig. 1.

{{Specifically, in our IC-NOSS scheme, SBS transmits a signal $x_S $ to SU with a normalized weight $w_S$, where $E(|{x_S}|^2 = 1)$. Meanwhile, MBS transmits its information-bearing signal $x_M $ at a power of $P_M-\bar P_m$ and a specially-designed signal $x_m $ at a power of $P_m$, where $E(|{x_M}|^2) =  E(|{x_m}|^2) = 1$ and $\bar P_m $ denotes an average power of the specially-designed signal $x_m$ in the range of $0 \le \bar P_m \le P_M$. In this way, the total average transmit power of $x_M$ and $x_m$ is constrained to $P_M$. Hence, we can express the received signal at MU as
\begin{equation}\label{equa1}
\begin{split}
  y_m^{{\text{IC}}} &= {h_{Mm}}( {\sqrt {{P_M} - {\bar P_m}} {x_M} + \sqrt{\bar P_m}{x_m}} )  + {h_{Sm}}\sqrt {{P_S}} w_S {x_S} + {n_m}, \\
\end{split}
\end{equation}
where $n_m$ represents an AWGN encountered at MU. In order to neutralize the interference received at MU, the following equation ${\sqrt{\bar P_m}}{h_{Mm}}{x_m} + \sqrt {{P_S}} {h_{Sm}}w_S{x_S} = 0$ should be satisfied, from which a special signal $x_m$ and its average power $\bar P_m$ as well as a normalized weight $w_S$ are given by
\begin{equation}\label{equa2}
x_m = - \frac{|{h_{Sm}}|}{{{\sigma _{Sm}}}} {e^{ - j{\theta _{Mm}}}}{x_S},
\end{equation}
and
\begin{equation}\label{equa3}
{\bar P_m} = \frac{{\sigma^2_{Sm}}}{{\sigma _{Mm}^2}}{P_S},
\end{equation}
and
\begin{equation}\label{equa4}
w_S = \frac{|{h_{Mm}}|}{{{\sigma _{Mm}}}}{e^{ - j{\theta _{Sm}}}},
\end{equation}
where ${\theta _{Mm}}$ and ${\theta _{Sm}}$ are the channel phases of MBS-MU and SBS-MU, respectively. One can observe from (2)-(4) that the eavesdropper's channel state information (CSI) of ${h_{Me}}$ and ${h_{Se}}$ is not needed, while
${h_{Mm}}$, ${h_{Sm}}$, $\sigma_{Mm}^2$, $\sigma_{Sm}^2$, $P_S$ and ${x_S}$ are required at MBS and SBS in carrying out the design of $[x_m, w_S]$. In general, the CSIs of $h_{Mm}$ and $h_{Sm}$ can be obtained at MU through channel estimation and then fed back to MBS and SBS over a multicast channel. The statistical information of $\sigma_{Mm}^2$ and $\sigma_{Sm}^2$ may be readily estimated by accumulating the instantaneous CSIs of $h_{Mm}$ and $h_{Sm}$. Moreover, MBS and SBS are often connected to the core network (e.g., a MSC in GSM, a MME in LTE, etc.) via fiber cables, through which the reliable information exchange of $x_S$ and $P_S$ can be achieved between MBS and SBS. Thus, both $x_S$ and $P_S$ can be acquired for designing $x_m$ at MBS through the core network. As discussed in [3], the message $x_S$ is typically initiated by another user terminal of cellular networks and sent via the core network first to SBS which then forwards to SU through its air interface. In other words, when the core network sends $x_S$ to SBS along with the power information $P_S$, the same copies of $x_S$ and $P_S$ can be received and stored at MBS simultaneously. This guarantees the synchronization between MBS and SBS and no significant amount of extra time delay incurred at MBS in obtaining $x_S$ and $P_S$ compared to SBS, regardless of the latency of the core network. It is of particular interest to examine the impact of channel estimation errors and feedback delay on the secrecy performance of our IC-NOSS scheme, which is considered for future work.}}

{{From (2) and (3), an instantaneous transmit power for the special signal $x_m$ is obtained as
\begin{equation}\label{equa5}
{ P_m} = \frac{{|h_{Sm}|^2}}{{\sigma _{Mm}^2}}{P_S}.
\end{equation}
Substituting $x_m$, $\bar P_m$ and $w_S$ from (2)-(4) into (1) yields
\begin{equation}\nonumber
y_m^{{\text{IC}}} = {h_{Mm}}\sqrt {{P_M} - {\bar P_m}} {x_M}   + {n_m},
\end{equation}
from which the MBS-MU channel capacity relying on our IC-NOSS scheme is given by
\begin{equation}\label{equa6}
C_{Mm}^{{\text{IC}}} = {\log _2}\left[{ 1 + ({\gamma _M} - \bar\gamma_m)|{h_{Mm}}{|^2} }\right],
\end{equation}
where $\gamma_M = P_M/N_0$ and $\bar\gamma_m = \bar P_m/N_0$. Let $\beta$ denote a ratio of the SNR of small cell to the SNR of macro cell (i.e., $ \beta = {\gamma _s}/{\gamma _M} $), called the small-to-macro ratio (SMR) for short, where $\gamma_S = P_S/N_0$. Noting that $\bar P_m$ as given by (3) should be in the range of $0 \le \bar P_m \le P_M$, we can obtain the following inequality
\begin{equation}\label{equa7}
0\le \beta \le \frac{{\sigma _{Mm}^2}}{{\sigma _{Sm}^2}},
\end{equation}
which should be satisfied such that the mutual interference can be completely canceled out at MU. Also, the received signal at SU can be written as
\begin{equation}\label{equa8}
y_s^{{\text{IC}}} = {h_{Ss}}\sqrt {{P_S}} w_S {x_S} + {h_{Ms}}( {\sqrt {{P_M} - {\bar P_m}} {x_M} + \sqrt {\bar P_m}{x_m}} ) + {n_s},
\end{equation}
where $w_S$, $\bar P_m$ and $x_m$ are given by (2)-(4), respectively. Treating $x_M$ and $x_m$ as interference, we can obtain the SBS-SU channel capacity as
\begin{equation}\label{equa9}
C_{Ss}^{{\text{IC}}} = {\log _2}[1 + \frac{{|{h_{Ss}}{|^2}{\gamma _S}{{\left| {{h_{Mm}}} \right|}^2}/\sigma^2_{Mm}}}{{|{h_{Ms}}{|^2}(\gamma _M - \bar\gamma_m + \gamma_m) + 1}}],
\end{equation}
where $\gamma_m = P_m/N_0$. Meanwhile, the signal transmissions of MBS and SBS may be overheard by the eavesdropper and the corresponding received signal is given by
\begin{equation}\label{equa10}
y_e^{{\text{IC}}} = {h_{Me}}( {\sqrt {{P_M} - {\bar P_m}} {x_M} + \sqrt{\bar P_m}{x_m}} )  + {h_{Se}}\sqrt {{P_S}} w_S {x_S} + {n_e},
\end{equation}
where $n_e$ represents an AWGN encountered at the eavesdropper. It can be seen from the preceding equation that the eavesdropper may decode the confidential signals of $x_M$ and $x_S$ with or without the successive interference cancelation (SIC). In this paper, we assume that no SIC is adopted by the eavesdropper for decoding $x_M$ and $x_S$. Although the SIC can be used for improving the achievable data rate of wiretap channel, it can also be employed by the legitimate SU for enhancing the data rate of SBS-SU transmissions. It can be expected that the use of SIC has a limited impact on the secrecy rate performance of our heterogeneous networks. Hence, by treating $x_m$ and $x_S$ as interference, the MBS-E channel capacity is obtained from (10) as
\begin{equation}\label{equa11}
C_{Me}^{{\text{IC}}} = {\log _2}[1 + \frac{{|{h_{Me}}{|^2}({\gamma _M} - \bar\gamma_m)}}{{|{h_{Me}}{|^2}\gamma _m + |{h_{Se}}{|^2}{\gamma _S}|{h_{Mm}}{|^2}/\sigma^2_{Mm} + 1}}].
\end{equation}
Similarly, the SBS-E channel capacity is given by
\begin{equation}\label{equa12}
C_{Se}^{{\text{IC}}} = {\log _2}[1 + \frac{{|{h_{Se}}{|^2}{\gamma _S}{{| {{h_{Mm}}} |}^2/\sigma^2_{Mm}}}}{|{h_{Me}}{|^2}(\gamma _M - \bar\gamma _m + \gamma _m) + 1}],
\end{equation}
which completes the signal model of IC-NOSS scheme.}}

\section{Secrecy Outage Probability Analysis}
In this section, we derive closed-form expressions for the secrecy outage probability of the OSS, IL-NOSS and IC-NOSS schemes over Rayleigh fading channels. As known, when the secrecy capacity of data transmissions denoted by $C_s$ falls below a predefined secrecy rate $R_s$, perfect secrecy is not achievable and a secrecy outage event is considered to occur in this case. Thus, the probability of occurrence of a secrecy outage event (referred to as secrecy outage probability) can be expressed as
\begin{equation}\label{equa13}
{{P_{\textrm{out} }} = \Pr \left( {{C_s} < {R_s}} \right),}
\end{equation}
where $C_s$ and $R_s$ represent the secrecy capacity and secrecy rate, respectively.

\subsection{OSS Scheme}
In this subsection, we analyze an overall secrecy outage probability of the macro-cell and small-cell transmissions relying on the OSS scheme, in which the macro cell and small cell take turns to access their shared spectrum. For notational convenience, let $\alpha$ denote a fraction of the total spectrum assigned to MBS and the remaining fraction $1-\alpha$ is allocated to SBS, where $0 \le \alpha \le 1$. Denoting the transmit powers of MBS and SBS by $P_M$ and $P_S$, respectively, we can obtain the channel capacity of macro-cell transmission from MBS to MU $C_{Mm}^{{\text{OSS}}}$ and that of small-cell transmission from SBS to SU $C_{Ss}^{{\text{OSS}}}$ as
\begin{equation}\nonumber
{C_{Mm}^{{\text{OSS}}} = \alpha {\log _2}( {1 + {\gamma _M}|{h_{Mm}}{|^2}} )},
\end{equation}
and
\begin{equation}\nonumber
{C_{Ss}^{{\text{OSS}}} = ( {1 - \alpha }){\log _2}( {1 + \gamma _S |{h_{Ss}}{|^2}}),}
\end{equation}
where $\gamma _M  = {P_M}/{N_0}$ and $\gamma_S  = {P_S}/{N_0}$ are referred to as signal-to-noise ratios (SNRs) of MBS and SBS, respectively, and $h_{Mm}$ and $h_{Ss}$ are fading coefficients of MBS-MU and SBS-SU channels, respectively. Also, the channel capacity of macro-cell wiretap link from MBS to E and that of small-cell wiretap link from SBS to E are given by
\begin{equation}\nonumber
{C_{Me}^{{\text{OSS}}} = \alpha {\log _2}( {1 + \gamma _M |{h_{Me}}{|^2}} ),}
\end{equation}
and
\begin{equation}\nonumber
{C_{Se}^{{\text{OSS}}} = ( {1 - \alpha } ){\log _2}( {1 + \gamma _S |{h_{Se}}{|^2}} ),}
\end{equation}
where $h_{Me}$ and $h_{Se}$ are fading coefficients of MBS-E and SBS-E channels, respectively. As a result, the secrecy capacity of macro-cell transmissions is given by the difference between the capacity of MBS-MU channel $C_{Mm}^{{\text{OSS}}}$ and that of MBS-E wiretap channel $C_{Me}^{{\text{OSS}}}$, namely
\begin{equation}\label{equa14}
C_{s,M}^{{\textrm{OSS}}} = \left(C_{Mm}^{{\textrm{OSS}}} - C_{Me}^{{\textrm{OSS}}}\right)^+,
\end{equation}
where $(x)^+ = \max(x,0)$. Similarly, the secrecy capacity of small-cell transmissions can be written as
\begin{equation}\label{equa15}
C_{s,S}^{{\textrm{OSS}}} = \left(C_{Ss}^{{\textrm{OSS}}} - C_{Se}^{{\textrm{OSS}}}\right)^+.
\end{equation}

Using (13) and (14), we can obtain the secrecy outage probability of macro-cell transmissions as
\begin{equation}\label{equa16}
{P_{{\textrm{out,}}M}^{{\textrm{OSS}}} = \Pr ( {C_{s,M}^{{\textrm{OSS}}} < R_M^s} ),}
\end{equation}
where $R_M^s$ denotes a predefined secrecy rate of the macro-cell transmission. Substituting $C_{s,M}^{{\textrm{OSS}}}$ from (14) into (16) yields
\begin{equation}\label{equa17}
{\begin{split}
P_{{\textrm{out}},M}^{{\textrm{OSS}}} &= \Pr [ {\alpha {{\log }_2}( {\frac{{1 + {\gamma _M}{{\left| {{h_{Mm}}} \right|}^2}}}{{1 + {\gamma _M}{{\left| {{h_{Me}}} \right|}^2}}}} ) < R_M^s} ]\\
 &= \Pr ( {{{\left| {{h_{Mm}}} \right|}^2} < \Lambda _M + {2^{\frac{{R_M^s}}{\alpha }}}{{\left| {{h_{Me}}} \right|}^2}} ),
\end{split}}
\end{equation}
where ${\Lambda _M} = ( {{2^{{{R_M^s}}/{\alpha }}} - 1} )/{\gamma _M}$. Noting that ${\left| {{h_{Mm}}} \right|^2}$ and ${\left| {{h_{Me}}} \right|^2}$ are independent exponentially distributed random variables with respective means of $\sigma _{Mm}^2$ and $\sigma _{Me}^2$, we have
\begin{equation}\label{equa18}
{P_{{\textrm{out,}}M}^{{\textrm{OSS}}} = 1 - \frac{{\sigma _{Mm}^2}}{{\sigma _{Mm}^2 + \sigma _{Me}^2( {{\Lambda _M}{\gamma _M} + 1} )}}\exp ( { - \frac{{{\Lambda _M}}}{{\sigma _{Mm}^2}}} ).}
\end{equation}
Similarly, by letting $R_S^s$ denote a secrecy rate of the small-cell transmission, the secrecy outage probability of the small-cell SBS-SU transmission relying on the OSS scheme can be obtained from (15) as
\begin{equation}\label{equa19}
{\begin{split}
P_{{\textrm{out}},S}^{{\textrm{OSS}}} &= \Pr ( {C_{s,S}^{{\textrm{OSS}}} < R_S^s} )= 1 - \frac{{\sigma _{Ss}^2}}{{\sigma _{Ss}^2 + \sigma _{Se}^2( {{\Lambda _S}{\gamma _S} + 1} )}}\exp ( { - \frac{{{\Lambda _S}}}{{\sigma _{Ss}^2}}} ),
\end{split}}
\end{equation}
where ${\Lambda _S} = [ {{2^{{{R_S^s}}/{(1-\alpha) }}} - 1} ]/{\gamma _S}$.

In order to show a joint effect of the individual secrecy outage probabilities of macro cell and small cell, we define an overall secrecy outage probability as the product of the secrecy outage probability of MBS-MU transmission and that of SBS-SU transmission. {{Besides the aforementioned multiplied secrecy outage probability definition, another overall secrecy outage probability in the normalized sum form may be given by a mean of the individual secrecy outage probabilities. It can be observed that the normalized sum secrecy outage probability would be dominated by the higher individual secrecy outage probability, especially when the lower individual secrecy outage probability is continuously improved to be sufficiently small even without degrading the higher individual one. By contrast, any improvements or degradations of both individual secrecy outage probabilities of the MBS-MU and SBS-SU can be readily reflected in the multiplied secrecy outage probability. Moreover, the IC-NOSS scheme is mainly designed to improve the MBS-MU secrecy without affecting the SBS-SU transmission, which motivates the use of the multiplied secrecy outage probability definition in this paper.}} As a consequence, an overall secrecy outage probability of the OSS scheme denoted by $P_{{\textrm{out}}}^{{\textrm{OSS}}}$ can be written as
\begin{equation}\label{equa20}
{P_{{\textrm{out}}}^{{\textrm{OSS}}} = P_{{\textrm{out}},M}^{{\textrm{OSS}}} \times P_{{\textrm{out}},S}^{{\textrm{OSS}}},}
\end{equation}
where $P_{{\textrm{out}},M}^{{\textrm{OSS}}}$ and $P_{{\textrm{out}},S}^{{\textrm{OSS}}}$ are given by (18) and (19), respectively.

\subsection{IL-NOSS Scheme}
This subsection presents the secrecy outage analysis of IL-NOSS scheme, where MBS and SBS are allowed to access their shared spectrum simultaneously and mutual interference occurs between the macro cell and small cell. To this end, certain power allocation may be employed for the sake of limiting such mutual interference below a tolerable level. Considering that MBS and SBS transmit their confidential information simultaneously over the same spectrum with respective powers of $P_M$ and $P_S$, we can obtain the MBS-MU and SBS-SU channel capacities for the IL-NOSS scheme as
\begin{equation}\nonumber
{C_{Mm}^{{\textrm{IL}}} = {\log _2}( {1 + \frac{{{\gamma _M}|{h_{Mm}}{|^2}}}{{{\gamma _S}|{h_{Sm}}{|^2} + 1}}} )},
\end{equation}
and
\begin{equation}\nonumber
{C_{Ss}^{{\textrm{IL}}} = {\log _2}( {1 + \frac{{{\gamma _S}|{h_{Ss}}{|^2}}}{{{\gamma _M}|{h_{Ms}}{|^2} + 1}}} )},
\end{equation}
where ${h_{Sm}}$ and ${h_{Ms}}$ are fading coefficients of SBS-MU and MBS-SU channels, respectively. Similarly, the corresponding MBS-E and SBS-E wiretap channel capacities for the IL-NOSS scheme, denoted by $C_{Me}^{{\textrm{IL}}}$ and $C_{Se}^{{\textrm{IL}}}$, respectively, are given by
\begin{equation}\nonumber
{C_{Me}^{{\textrm{IL}}} = {\log _2}( {1 + \frac{{{\gamma _M}|{h_{Me}}{|^2}}}{{{\gamma _S}|{h_{Se}}{|^2} + 1}}}),}
\end{equation}
and
\begin{equation}\nonumber
{C_{Se}^{{\textrm{IL}}} = {\log _2}( {1 + \frac{{{\gamma _S}|{h_{Se}}{|^2}}}{{{\gamma _M}|{h_{Me}}{|^2} + 1}}} ).}
\end{equation}
As a consequence, the secrecy capacity of macro-cell and small-cell transmissions relying on the IL-NOSS scheme can be respectively expressed as
\begin{equation}\label{equa21}
C_{s,M}^{{\textrm{IL}}} = \left(C_{Mm}^{{\textrm{IL}}} - C_{Me}^{{\textrm{IL}}}\right)^+,
\end{equation}
and
\begin{equation}\label{equa22}
C_{s,S}^{{\textrm{IL}}} = \left(C_{Ss}^{{\textrm{IL}}} - C_{Se}^{{\textrm{IL}}}\right)^+,
\end{equation}
where $C_{Mm}^{{\textrm{IL}}}$, $C_{Me}^{{\textrm{IL}}}$, $C_{Ss}^{{\textrm{IL}}}$ and $C_{Se}^{{\textrm{IL}}}$ are the MBS-MU, MBS-E, SBS-SU and SBS-E channel capacities of IL-NOSS scheme, respectively.

\begin{table}
  \centering
  \textbf{\caption{}\label{table1}}
  \begin{tabular}{|c|c|}
  \hline
\textbf{\textbf{Parameters}}& \textbf{\textbf{Values}}\\
\hline
\textbf{\textbf{${a_1}$}}& \textbf{\textbf{$1/\sigma _{Mm}^2$}} \\
\hline
\textbf{\textbf{${b_1}$}}& \textbf{\textbf{${\gamma _S}\sigma _{Sm}^2/\sigma _{Mm}^2$}}\\
\hline
\textbf{\textbf{${c_1}$}}& \textbf{\textbf{$1/\sigma _{Me}^2$}}\\
\hline
\textbf{\textbf{${d_1}$}}& \textbf{\textbf{${{\gamma _S}\sigma _{Se}^2/\sigma _{Me}^2}$}}\\
\hline
\textbf{\textbf{${e_1}$}}& \textbf{\textbf{${\Delta _M}$}}\\
\hline
\textbf{\textbf{${f_1}$}}& \textbf{\textbf{${\gamma _M}{\Delta _M} + 1$}}\\
\hline
\textbf{\textbf{${g_1}$}}& \textbf{\textbf{$[{b_1}{f_1}({c_1}+d_1) - {c_1}d_1(b_1e_1+1)]/{[{b_1}{f_1} - {d_1}({b_1}{e_1} + 1)]^2}$}}\\
\hline
\textbf{\textbf{${h_1}$}}& \textbf{\textbf{$({b_1}{e_1} + 1)({a_1}{f_1} + {c_1})/({b_1}{f_1})$}}\\
\hline
\textbf{\textbf{${i_1}$}}&\textbf{\textbf{$({a_1}{f_1} + {c_1})/{d_1}$}}\\
\hline
\textbf{\textbf{${j_1}$}}& \textbf{\textbf{$ - {d_1}/[{b_1}{f_1} - {d_1}({b_1}{e_1} + 1)]$}}\\
\hline
\textbf{\textbf{${k_1}$}}& \textbf{\textbf{$[{a_1}{b_1}({d_1}{e_1} - {f_1}) + ({a_1} + {b_1}){d_1}]{f_1}/{[{b_1}{f_1} - {d_1}({b_1}{e_1} + 1)]^2}$}}\\
\hline
\textbf{\textbf{${l_1}$}}& \textbf{\textbf{$({c_1} - {a_1}{f_1})/[2{d_1}({b_1}{e_1} + 1)]$}}\\
\hline
  \end{tabular}
\end{table}

From (13) and (21), the secrecy outage probability of MBS-MU transmission relying on the IL-NOSS approach can be obtained as
\begin{equation}\label{equa23}
{P_{{\textrm{out}},M}^{{\textrm{IL}}} = \Pr ( {C_{s,M}^{{\textrm{IL}}} < R_M^s} ).}
\end{equation}
Substituting $C_{s,M}^{{\textrm{IL}}}$ from (21) into (23), we have
\begin{equation}\label{equa24}
{\begin{split}
P_{{\textrm{out}},M}^{{\textrm{IL}}}  = \Pr \left[ {\frac{{{{\left| {{h_{Mm}}} \right|}^2}}}{{{\gamma _S}{{\left| {{h_{Sm}}} \right|}^2} + 1}} < {\Delta _M} + \frac{{{{\left| {{h_{Me}}} \right|}^2( {{\gamma _M}{\Delta _M} + 1} )}}}{{{\gamma _S}{{\left| {{h_{Se}}} \right|}^2} + 1}}} \right],
\end{split}}
\end{equation}
where ${\Delta _M} = \left( {{2^{R_M^s}} - 1} \right)/{\gamma _M}$. Using Appendix A, we can obtain a closed-form expression for $P_{{\textrm{out}},M}^{{\textrm{IL}}} $ as
\begin{equation}\label{equa25}
{P_{{\textrm{out}},M}^{{\textrm{IL}}} = \left\{ \begin{split}
&1-\exp ( -{{a}_{1}}{{e}_{1}} )\times\left(
\begin{split}
&-g_1\exp(h_1)Ei(-h_1)\\
&+k_1\exp(i_1)Ei(-i_1)+j_1
\end{split}
\right)
,&b_1f_1 \ne (b_1e_1+1){d_1}\\
&1 - \exp ( { - {a_1}{e_1}} )[ l_1+\frac{d_1}{2b_1f_1}+l_1i_1\exp(i_1)Ei(-i_1)]
,&b_1f_1 = (b_1e_1+1){d_1}
\end{split} \right.},
\end{equation}
where $Ei(x ) = \int_x^{ \infty }{\frac{{{e^{-t}}}}{t}dt} $ is known as the exponential integral function and other used parameters are specified in Table I.

\begin{table}
  \centering
  \textbf{\caption{}\label{table2}}
  \begin{tabular}{|c|c|}
  \hline
\textbf{\textbf{Parameters}}& \textbf{\textbf{Values}}\\
\hline
\textbf{\textbf{${a_2}$}}& \textbf{\textbf{$1/\sigma _{Ss}^2$}} \\
\hline
\textbf{\textbf{${b_2}$}}& \textbf{\textbf{${\gamma _M}\sigma _{Ms}^2/\sigma _{Ss}^2$}}\\
\hline
\textbf{\textbf{${c_2}$}}& \textbf{\textbf{$1/\sigma _{Se}^2$}}\\
\hline
\textbf{\textbf{${d_2}$}}& \textbf{\textbf{${{\gamma _M}\sigma _{Me}^2/\sigma _{Se}^2}$}}\\
\hline
\textbf{\textbf{${e_2}$}}& \textbf{\textbf{${\Delta _S}$}}\\
\hline
\textbf{\textbf{${f_2}$}}& \textbf{\textbf{${\gamma _S}{\Delta _S} + 1$}}\\
\hline
\textbf{\textbf{${g_2}$}}& \textbf{\textbf{$[{b_2}{f_2}({c_2}+d_2) - {c_2}d_2(b_2e_2+1)]/{[{b_2}{f_2} - {d_2}({b_2}{e_2} + 1)]^2}$}}\\
\hline
\textbf{\textbf{${h_2}$}}& \textbf{\textbf{$({b_2}{e_2} + 1)({a_2}{f_2} + {c_2})/({b_2}{f_2})$}}\\
\hline
\textbf{\textbf{${i_2}$}}&\textbf{\textbf{$({a_2}{f_2} + {c_2})/{d_2}$}}\\
\hline
\textbf{\textbf{${j_2}$}}& \textbf{\textbf{$ - {d_2}/[{b_2}{f_2} - {d_2}({b_2}{e_2} + 1)]$}}\\
\hline
\textbf{\textbf{${k_2}$}}& \textbf{\textbf{$[{a_2}{b_2}({d_2}{e_2} - {f_2}) + ({a_2} + {b_2}){d_2}]{f_2}/{[{b_2}{f_2} - {d_2}({b_2}{e_2} + 1)]^2}$}}\\
\hline
\textbf{\textbf{${l_2}$}}& \textbf{\textbf{$({c_2} - {a_2}{f_2})/[2{d_2}({b_2}{e_2} + 1)]$}}\\
\hline
  \end{tabular}
\end{table}

Similarly, the secrecy outage probability of small-cell transmission can be obtained from (22) as
\begin{equation}\label{equa26}
{\begin{split}
&P_{{\textrm{out}},S}^{{\textrm{IL}}} = \Pr ( {C_{s,S}^{{\textrm{IL}}} < R_S^s} )= \Pr \left[ {\frac{{{{\left| {{h_{Ss}}} \right|}^2}}}{{{\gamma _M}{{\left| {{h_{Ms}}} \right|}^2} + 1}} < {\Delta _S} + \frac{{{{\left| {{h_{Se}}} \right|}^2( {{\gamma _S}{\Delta _S} + 1} )}}}{{{\gamma _M}{{\left| {{h_{Me}}} \right|}^2} + 1}}} \right],
\end{split}}
\end{equation}
which can be further given by
\begin{equation}\label{equa27}
{P_{{\textrm{out}},S}^{{\textrm{IL}}} = \left\{ \begin{split}
&1-\exp ( -{{a}_{2}}{{e}_{2}} )\times\left(
\begin{split}
&-g_2\exp(h_2)Ei(-h_2)\\
&+k_2\exp(i_2)Ei(-i_2)+j_2
\end{split}
\right)
,&b_2f_2 \ne (b_2e_2+1){d_2}\\
&1 - \exp ( { - {a_2}{e_2}} )[ l_2+\frac{d_2}{2b_2f_2}+l_2i_2\exp(i_2)Ei(-i_2)]
,&b_2f_2 = (b_2e_2+1){d_2}
\end{split} \right.},
\end{equation}
where the used parameters are stated in Table II. Similar to (20), an overall secrecy outage probability of the IL-NOSS scheme is obtained as
\begin{equation}\label{equa28}
{P_{{\textrm{out}}}^{{\textrm{IL}}} = P_{{\textrm{out}},M}^{{\textrm{IL}}} \times P_{{\textrm{out}},S}^{{\textrm{IL}}},}
\end{equation}
where $P_{{\textrm{out}},M}^{{\textrm{IL}}}$ and $P_{{\textrm{out}},S}^{{\textrm{IL}}}$ are the individual secrecy outage probabilities of the macro cell and small cell as given by (25) and (27), respectively.

\subsection{IC-NOSS Scheme}
{{In this subsection, we present the secrecy outage probability analysis of IC-NOSS scheme. Using (6), (9), (11) and (12), we can obtain the secrecy capacity of MBS-MU and SBS-SU transmissions relying on the IC-NOSS scheme as
\begin{equation}\label{equa29}
C_{{s,}M}^{{\textrm{IC}}} = \left(C_{Mm}^{{\textrm{IC}}} - C_{Me}^{{\textrm{IC}}}\right)^+
\end{equation}
and
\begin{equation}\label{equa30}
C_{{s,}S}^{{\textrm{IC}}} = \left(C_{Ss}^{{\textrm{IC}}} - C_{Se}^{{\textrm{IC}}}\right)^+,
\end{equation}
where $C_{Mm}^{{\textrm{IC}}}$, $C_{Me}^{{\textrm{IC}}}$, $C_{Ss}^{{\textrm{IC}}}$ and $C_{Se}^{{\textrm{IC}}}$ are the MBS-MU, MBS-E, SBS-SU and SBS-E channel capacities for the IC-NOSS scheme, respectively.}}

{{Using (13) and (29), we can obtain the secrecy outage probability of macro cell for the IC-NOSS scheme as
\begin{equation}\label{equa31}
P_{{\textrm{out,}}M}^{{\textrm{IC}}} = \Pr \left[ {\dfrac{{1 + ({\gamma _M} - {{\bar \gamma }_m})|{h_{Mm}}{|^2}}}{{1 + \dfrac{{|{h_{Me}}{|^2}({\gamma _M} - {{\bar \gamma }_m})}}{{|{h_{Me}}{|^2}{\gamma _m} + |{h_{Se}}{|^2}|{h_{Mm}}{|^2}\sigma _{Mm}^{ - 2}{\gamma _S} + 1}}}} < {2^{R_M^s}}} \right],
\end{equation}
which is further given by
\begin{equation}\label{equa32}
P_{{\textrm{out,}}M}^{{\textrm{IC}}} = \Pr \left[ \begin{split}
 &\left( {\frac{{1 - {2^{R_M^s}}}}{{{\gamma _M}}} + \frac{{|{h_{Mm}}{|^2}({\gamma _M} - {{\bar \gamma }_m})}}{{{\gamma _M}}}} \right) \\
  &\times \left( {\frac{{{\gamma _m}}}{{{\gamma _M}}} + \frac{{|{h_{Se}}{|^2}|{h_{Mm}}{|^2}{\gamma _S}}}{{\sigma _{Mm}^2|{h_{Me}}{|^2}{\gamma _M}}} + \frac{1}{{{\gamma _M}}}} \right) \\
  &< \frac{{{2^{R_M^s}}({\gamma _M} - {{\bar \gamma }_m})}}{{\gamma _M^2}} \\
 \end{split} \right].
\end{equation}}}

{{It can be observed from (32) that deriving a general closed-form expression for $P_{{\textrm{out,}}M}^{{\textrm{IC}}}$ is challenging. As a consequence, let us consider an asymptotic case with $\gamma_M \to \infty$, for which the terms of ${\frac{{1 - {2^{R_M^s}}}}{{{\gamma _M}}}}$ and ${\frac{1}{{{\gamma _M}}}}$ of (32) are high order infinitesimals and negligible, leading to
\begin{equation}\label{equa33}
P_{{\textrm{out,}}M}^{{\textrm{IC}}} =\Pr \left[ {|{h_{Mm}}{|^2}\left( {|{h_{Sm}}{|^2} + \frac{{|{h_{Se}}{|^2}|{h_{Mm}}{|^2}}}{{|{h_{Me}}{|^2}}}} \right) < \frac{{\sigma _{Mm}^2{2^{R_M^s}}}}{{{\beta\gamma _M}}}} \right],
\end{equation}
where $\beta = \gamma_S/\gamma_M$. From Appendix B, we obtain
\begin{equation}\label{equa34}
P_{{\textrm{out,}}M}^{{\textrm{IC}}} = \int_0^\infty  {f(x)\exp ( - x)dx},
\end{equation}
where $f(x)$ is given by
\begin{equation}\nonumber
f(x)  = \frac{{{2^{R_M^s}}}}{{\sigma _{Sm}^2\beta {\gamma _M}{\varphi _x}x}} - {\phi _x}\exp ( - {\varphi _x})\int_{ - {\varphi _x}}^{ - {\phi _x}} {\frac{1}{{{t^2}}}\exp ( - t)dt} ,
\end{equation}
wherein ${\varphi_x} = \frac{{\sigma _{Mm}^2\sigma _{Se}^2\beta {\gamma _M}{x^2} + \sigma _{Me}^2{2^{R_M^s}}}}{{\sigma _{Sm}^2\sigma _{Me}^2\beta {\gamma _M}x}}$ and ${\phi_x}  = \frac{{\sigma _{Mm}^2\sigma _{Se}^2x}}{{\sigma _{Sm}^2\sigma _{Me}^2}}$. Similarly, the secrecy outage probability of small-cell transmission relying on the IC-NOSS scheme can be obtained from (30) as
\begin{equation}\label{equa35}
P_{{\textrm{out,}}S}^{{\textrm{IC}}} = \Pr \left[
 \dfrac{{|{h_{Ss}}{|^2}|{h_{Mm}}{|^2}/\sigma _{Mm}^2}}{{|{h_{Ms}}{|^2}({\gamma _M} - {{\bar \gamma }_m} + {\gamma _m}) + 1}} <{\Delta _S} + \dfrac{{{2^{R_S^s}}|{h_{Se}}{|^2}|{h_{Mm}}{|^2}/\sigma _{Mm}^2}}{{|{h_{Me}}{|^2}({\gamma _M} - {{\bar \gamma }_m} + {\gamma _m}) + 1}} \\
\right],
\end{equation}
wherein ${\Delta _S} = ({{{2^{R_S^s}} - 1}})/{{{\gamma _S}}}$. It is challenging to derive a general closed-form expression for $P_{{\textrm{out,}}S}^{{\textrm{IC}}}$ from the preceding equation of (35). To this end, we consider a special case with a sufficiently small interference gain of $\sigma^2_{Sm} \ll 1$, which is possible in practical heterogeneous cellular systems, since a small cell may be deployed in some shadowed areas of the macro cell e.g. in-building places, underground garages, tunnels [1]. Hence, one can obtain from (3) and (5) that both ${{ \gamma }_m} $ and ${{\bar \gamma }_m} $ are negligible compared to $\gamma_M$ for the case of $\sigma^2_{Sm} \ll 1$, from which (35) is given by
\begin{equation}\label{equa36}
P_{{\textrm{out,}}S}^{{\textrm{IC}}} = \Pr \left[ {\frac{{|{h_{Ss}}{|^2}|{h_{Mm}}{|^2}/\sigma _{Mm}^2}}{{|{h_{Ms}}{|^2}{\gamma _M} + 1}} < {\Delta _S} + \frac{{{2^{R_S^s}}|{h_{Se}}{|^2}|{h_{Mm}}{|^2}/\sigma _{Mm}^2}}{{|{h_{Me}}{|^2}{\gamma _M} + 1}}} \right].
\end{equation}
Noting $|{h_{Mm}}{|^2}$, $|{h_{Ss}}{|^2}$, $|{h_{Ms}}{|^2}$, $|{h_{Me}}{|^2}$ and $|{h_{Se}}{|^2}$ are independent exponentially distributed random variables with respective means of $\sigma^2_{Mm}$, $\sigma^2_{Ss}$, $\sigma^2_{Ms}$, $\sigma^2_{Me}$ and $\sigma^2_{Se}$, we have
\begin{equation}\label{equa37}
P_{{\textrm{out,}}S}^{{\textrm{IC}}} = \int_0^\infty  {g(x)\exp ( - x)} dx,
\end{equation}
where $g(x)$ is given by
\begin{equation}\label{equa38}
g(x) = \Pr \left[ {\frac{{|{h_{Ss}}{|^2}x}}{{|{h_{Ms}}{|^2}{\gamma _M} + 1}} < {\Delta _S} + \frac{{|{h_{Se}}{|^2}x({\Delta _S}{\gamma _S} + 1)}}{{|{h_{Me}}{|^2}{\gamma _M} + 1}}} \right].
\end{equation}}}

\begin{table}
  \centering
  \textbf{\caption{}\label{table2}}
  \begin{tabular}{|c|c|}
  \hline
\textbf{\textbf{Parameters}}& \textbf{\textbf{Values}}\\
\hline
\textbf{\textbf{${a_x}$}}& \textbf{\textbf{$1/(\sigma _{Ss}^2x)$}} \\
\hline
\textbf{\textbf{${b_x}$}}& \textbf{\textbf{${\gamma _M}\sigma _{Ms}^2/(\sigma _{Ss}^2x)$}}\\
\hline
\textbf{\textbf{${c_x}$}}& \textbf{\textbf{$1/(\sigma _{Se}^2x)$}}\\
\hline
\textbf{\textbf{${d_x}$}}& \textbf{\textbf{${{\gamma _M}\sigma _{Me}^2/(\sigma _{Se}^2x)}$}}\\
\hline
\textbf{\textbf{${e_x}$}}& \textbf{\textbf{${\Delta _S}$}}\\
\hline
\textbf{\textbf{${f_x}$}}& \textbf{\textbf{${\gamma _S}{\Delta _S} + 1$}}\\
\hline
\textbf{\textbf{${g_x}$}}& \textbf{\textbf{$[{b_x}{f_x}({c_x}+d_x) - {c_x}d_x(b_xe_x+1)]/{[{b_x}{f_x} - {d_x}({b_x}{e_x} + 1)]^2}$}}\\
\hline
\textbf{\textbf{${h_x}$}}& \textbf{\textbf{$({b_x}{e_x} + 1)({a_x}{f_x} + {c_x})/({b_x}{f_x})$}}\\
\hline
\textbf{\textbf{${i_x}$}}&\textbf{\textbf{$({a_x}{f_x} + {c_x})/{d_x}$}}\\
\hline
\textbf{\textbf{${j_x}$}}& \textbf{\textbf{$ - {d_x}/[{b_x}{f_x} - {d_x}({b_x}{e_x} + 1)]$}}\\
\hline
\textbf{\textbf{${k_x}$}}& \textbf{\textbf{$[{a_x}{b_x}({d_x}{e_x} - {f_x}) + ({a_x} + {b_x}){d_x}]{f_x}/{[{b_x}{f_x} - {d_x}({b_x}{e_x} + 1)]^2}$}}\\
\hline
\textbf{\textbf{${l_x}$}}& \textbf{\textbf{$({c_x} - {a_x}{f_x})/[2{d_x}({b_x}{e_x} + 1)]$}}\\
\hline
  \end{tabular}
\end{table}

{{As observed from (26) and (38), one can readily obtain $g(x)$ as
\begin{equation}\label{equa39}
g(x) = \left\{ \begin{split}
&1-\exp ( -{{a}_{x}}{{e}_{x}} )\times\left(
\begin{split}
&-g_x\exp(h_x)Ei(-h_x)\\
&+k_x\exp(i_x)Ei(-i_x)+j_x
\end{split}
\right)
,&b_xf_x \ne (b_xe_x+1){d_x}\\
&1 - \exp ( { - {a_x}{e_x}} )[ l_x+\frac{d_x}{2b_xf_x}+l_xi_x\exp(i_x)Ei(-i_x)]
,&b_xf_x = (b_xe_x+1){d_x}
\end{split} \right.,
\end{equation}
where the parameters of $a_x$, $b_x$, $c_x$, $d_x$, $e_x$, $f_x$, $g_x$, $h_x$, $i_x$, $k_x$ and $l_x$ are specified in Table III. Finally, an overall secrecy outage probability of the IC-NOSS scheme is expressed as
\begin{equation}\label{equa40}
{P_{{\textrm{out}}}^{{\textrm{IC}}} = P_{{\textrm{out}},M}^{{\textrm{IC}}} \times P_{{\textrm{out}},S}^{{\textrm{IC}}},}
\end{equation}
where $P_{{\textrm{out}},M}^{{\textrm{IC}}}$ and $P_{{\textrm{out}},S}^{{\textrm{IC}}}$ are given by (34) and (37), respectively.}}

\section{Secrecy Diversity Gain Analysis}
In this section, we present the secrecy diversity analysis of OSS, IL-NOSS and IC-NOSS schemes. The secrecy diversity is used to characterize an asymptotic behavior of the secrecy outage probability in the high SNR region, which is defined as a ratio of the logarithm of secrecy outage probability to that of SNR, as the SNR approaches to infinity, namely
\begin{equation}\label{equa41}
{{d_s} =  - \mathop {\lim }\limits_{{\gamma _M} \to \infty } \frac{{\log {P_{{\textrm{out}}}}}}{{\log {\gamma _M}}},}
\end{equation}
where ${P_{{\textrm{out}}}}$ denotes a secrecy outage probability and $\gamma_M$ is an SNR.

\subsection{OSS Scheme}
This subsection analyzes the secrecy diversity gain of OSS scheme, which is given by
\begin{equation}\label{equa42}
{d_{s}^{{\textrm{OSS}}} =  - \mathop {\lim }\limits_{{\gamma _M} \to \infty } \frac{{\log P_{{\textrm{out}}}^{{\textrm{OSS}}}}}{{\log {\gamma _M}}},}
\end{equation}
where $P_{{\textrm{out}}}^{{\textrm{OSS}}}$ is the overall secrecy outage probability of OSS scheme as given by (20). Substituting $P_{{\textrm{out}}}^{{\textrm{OSS}}}$ from (20) into (42) yields
\begin{equation}\label{equa43}
{d_s^{{\textrm{OSS}}} =  - \mathop {\lim }\limits_{{\gamma _M} \to \infty } \frac{{\log (P_{{\textrm{out}},M}^{{\textrm{OSS}}} \cdot P_{{\textrm{out}},S}^{{\textrm{OSS}}})}}{{\log {\gamma _M}}}}.
\end{equation}
Denoting ${\gamma _S} = \beta {\gamma _M}$ and substituting $P_{{\textrm{out}},M}^{{\textrm{OSS}}}$ and $P_{{\textrm{out}},S}^{{\textrm{OSS}}}$ from (18) and (19) into (43), we can readily arrive at
\begin{equation}\label{equa44}
{d_{s}^{{\textrm{OSS}}} = 0,}
\end{equation}
which shows that a secrecy diversity gain of zero is achieved by the OSS scheme. This means that as the SNR increases to infinity, the overall secrecy outage probability of OSS scheme converges to a secrecy outage floor and would not be arbitrarily low.

\subsection{IL-NOSS Scheme}
In this subsection, we present the secrecy diversity analysis of IL-NOSS scheme. Similar to (39), we can obtain the secrecy diversity of IL-NOSS scheme as
\begin{equation}\label{equa45}
{d_s^{{\textrm{IL}}} =  - \mathop {\lim }\limits_{{\gamma _M} \to \infty } \frac{{\log (P_{{\textrm{out}},M}^{{\textrm{IL}}} \cdot P_{{\textrm{out}},S}^{{\textrm{IL}}})}}{{\log {\gamma _M}}}},
\end{equation}
where $P_{{\textrm{out}},M}^{{\textrm{IL}}}$ and $P_{{\textrm{out}},S}^{{\textrm{IL}}}$ are given by (24) and (26), respectively. Denoting ${\gamma _S} = \beta {\gamma _M}$ and letting ${\gamma _M} \to \infty $, we can obtain the secrecy outage probability of MBS-MU transmissions relying on the IL-NOSS scheme from (24) as
\begin{equation}\label{equa46}
\mathop {\lim }\limits_{{\gamma _M} \to \infty } {P_{{\textrm{out},}M}^{{\textrm{IL}}} = \Pr \left( {\frac{{{{\left| {{h_{Mm}}} \right|}^2}}}{{{{\left| {{h_{Sm}}} \right|}^2}}} <  \frac{{{{{2^{R_M^s}}}{\left| {{h_{Me}}} \right|}^2} }}{{{{\left| {{h_{Se}}} \right|}^2}}}} \right)},
\end{equation}
which is obtained by ignoring the background noise in the high SNR region. For national convenience, let $X = \frac{{{{\left| {{h_{Mm}}} \right|}^2}}}{{{{\left| {{h_{Sm}}} \right|}^2}}}$ and $Y = \frac{{{{\left| {{h_{Me}}} \right|}^2}}}{{{{\left| {{h_{Se}}} \right|}^2}}}$. Since ${\left| {{h_{Mm}}} \right|^2}$, ${\left| {{h_{Sm}}} \right|^2}$, ${\left| {{h_{Me}}} \right|^2}$ and ${\left| {{h_{Se}}} \right|^2}$ are independent exponentially distributed random variables with respective means of $\sigma _{Mm}^2$, $\sigma _{Sm}^2$, $\sigma _{Me}^2$ and $\sigma _{Se}^2$, the cumulative distribution functions (CDFs) of $X$ and $Y$ are obtained as
\begin{equation}\label{equa47}
\Pr\left( X<x \right) = 1 - \frac{{\sigma _{Mm}^2}}{{\sigma _{Mm}^2 + \sigma _{Sm}^2x}},
\end{equation}
and
\begin{equation}\label{equa48}
\Pr\left( Y<y \right)  = 1 - \frac{{\sigma _{Me}^2}}{{\sigma _{Me}^2 + \sigma _{Se}^2y}},
\end{equation}
from which the probability density function (PDF) of $Y$ is given by
\begin{equation}\label{equa49}
{p_Y}\left( y \right) = \frac{{\sigma _{Se}^2\sigma _{Me}^2}}{{{{( {\sigma _{Me}^2 + \sigma _{Se}^2y} )}^2}}}.
\end{equation}
By using (47) and (49), (46) is obtained as
\begin{equation}\label{equa50}
\mathop {\lim }\limits_{{\gamma _M} \to \infty } P_{{\textrm{out},}M}^{{\textrm{IL}}}
= \int_0^\infty  {\left( {1 - \frac{{\sigma _{Mm}^2}}{{\sigma _{Sm}^2 2^{R_M^s}y+ \sigma _{Mm}^2}}} \right)} \frac{{\sigma _{Se}^2\sigma _{Me}^2}}{{{{( { \sigma _{Me}^2+\sigma _{Se}^2y } )}^2}}}dy,
\end{equation}
which converges to a non-zero secrecy outage probability floor, implying
\begin{equation}\label{equa51}
\mathop {\lim }\limits_{{\gamma _M} \to \infty } \frac{{\log (P_{{\textrm{out}},M}^{{\textrm{IL}}} )}}{{\log {\gamma _M}}} = 0.
\end{equation}
Similarly, a secrecy outage probability floor also happens for the small-cell transmission for $\gamma_M \to \infty$, namely
\begin{equation}\label{equa52}
\mathop {\lim }\limits_{{\gamma _M} \to \infty } \frac{{\log (P_{{\textrm{out}},S}^{{\textrm{IL}}} )}}{{\log {\gamma _M}}} = 0.
\end{equation}
As a consequence, by substituting (51) and (52) into (45), the secrecy diversity gain of the IL-NOSS scheme can be readily obtained as
\begin{equation}\renewcommand\label{equa53}
{d_{s}^{{\textrm{IL}}} = 0,}
\end{equation}
which implies that the overall secrecy outage probability of IL-NOSS scheme would not decrease to be arbitrarily small, as the SNR $\gamma_M$ increases to infinity.

\subsection{IC-NOSS scheme}
{{This subsection presents the secrecy diversity analysis of IC-NOSS scheme. Similar to (45), we can obtain the secrecy diversity of IC-NOSS scheme as
\begin{equation}\label{equa54}
{d_s^{{\textrm{IC}}} =  - \mathop {\lim }\limits_{{\gamma _M} \to \infty } \frac{{\log (P_{{\textrm{out}},M}^{{\textrm{IC}}} \cdot P_{{\textrm{out}},S}^{{\textrm{IC}}})}}{{\log {\gamma _M}}}},
\end{equation}
where $P_{{\textrm{out}},M}^{{\textrm{IC}}}$ and $P_{{\textrm{out}},S}^{{\textrm{IC}}}$ are given by (33) and (35), respectively. From Appendix C, we can obtain lower and upper bounds of $P_{{\textrm{out}},M}^{{\textrm{IC}}}$ as
\begin{equation}\label{equa55}
P_{{\textrm{out,}}M}^{{\textrm{IC}}} \ge \frac{{{2^{R_M^s}}\ln ({\gamma _M})}}{{8\sigma _{Sm}^2\beta {\gamma _M}}},
\end{equation}
and
\begin{equation}\label{equa56}
P_{{\textrm{out,}}M}^{{\textrm{IC}}} \le \frac{{{2^{R_M^s}}\ln ({\gamma _M})}}{{2\sigma _{Sm}^2\beta {\gamma _M}}},
\end{equation}
for $\gamma_M \to \infty$.}}

\begin{figure}
\centering
\includegraphics[scale = 0.75]{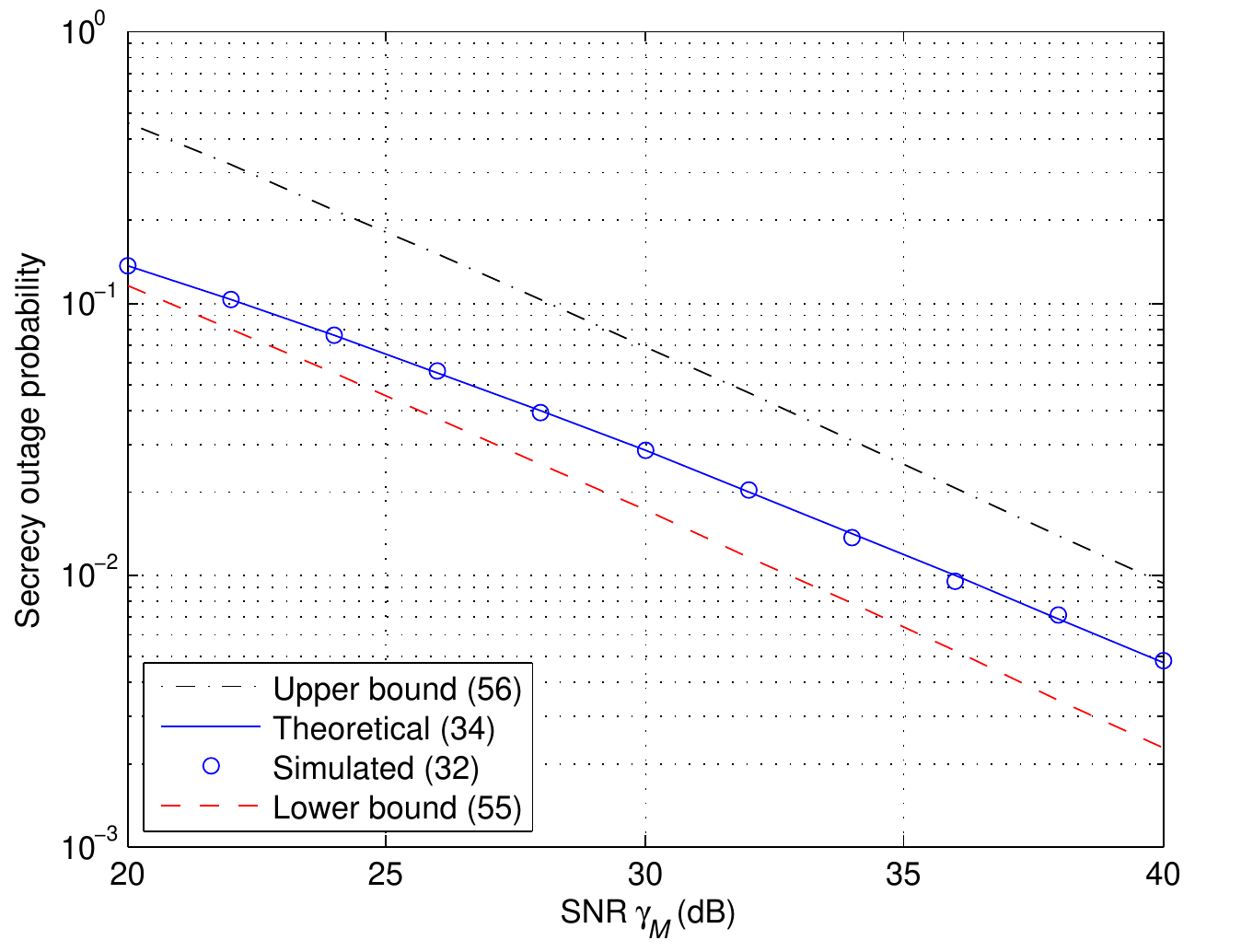}
\caption{Comparisons among the secrecy outage probability of macro-cell transmissions $P_{{\textrm{out,}}M}^{{\textrm{IC}}}$ as well as its lower and upper bounds versus SNR $\gamma_M$ for the IC-NOSS scheme with $R_M^s =1{\textrm{bit/s/Hz}}$, $\sigma _{Mm}^2 = \sigma _{Me}^2 = \sigma _{Se}^2 = 1$, $\sigma _{Sm}^2 = 0.2$, $\alpha  = 0.5$, and $\beta = 0.5 $.}
\label{fig 2}
\end{figure}

{{To verify the effectiveness of our derived lower and upper bounds of the secrecy outage probability as given by (55) and (56), we show some numerical secrecy outage comparisons of macro-cell transmissions for the IC-NOSS scheme in Fig. 2. It is pointed out that the theoretical secrecy outage probability curve of Fig. 2 is plotted by using (34) and the corresponding simulated result is obtained through the Monte-Carlo simulation of (32). Moreover, in Fig. 2, a smaller fading gain of $ \sigma _{Sm}^2 = 0.2$ is considered for an interference channel from a small cell to macro cell, which is due to the fact that the small cell is typically deployed in a shadowed area e.g. in-building places of the macro cell. One can observe from Fig. 2 that the theoretical and simulated secrecy outage probabilities match well with each other, demonstrating the correctness of our secrecy outage analysis. Fig. 2 also shows that both the theoretical and simulated secrecy outage probability results fall between the lower and upper bounds in the high SNR region, validating the effectiveness of our derived lower and upper bounds of (55) and (56) for $\gamma_M \to \infty$.}}

{{As implied from (55) and (56), the secrecy outage probability of macro-cell transmission for the IC-NOSS scheme $P_{\text{zero,}M}^{\text{IC}}$ behaves as $\frac{\ln(\gamma_M)}{{{\gamma _M}}}$, as the SNR $\gamma_M$ increases to infinity, namely
\begin{equation}\nonumber
-1 \le \mathop {\lim }\limits_{{\gamma _M} \to \infty } \frac{ \log P_{{\textrm{out,}}M}^{{\textrm{IC}}} }{{\log {\gamma _M}}} \le  - 1,
\end{equation}
which leads to
\begin{equation}\label{equa57}
\mathop {\lim }\limits_{{\gamma _M} \to \infty } \frac{ \log P_{{\textrm{out,}}M}^{{\textrm{IC}}} }{{\log {\gamma _M}}} = - 1.
\end{equation}
Additionally, one can observe from (35) that $P_{{\textrm{out},}S}^{{\textrm{IC}}}$ converges to a secrecy outage probability floor for $\gamma_M \to \infty$, implying
\begin{equation}\label{equa58}
\mathop {\lim }\limits_{{\gamma _M} \to \infty } \frac{{\log P_{{\textrm{out,}}S}^{{\textrm{IC}}}}}{{\log {\gamma _M}}} = 0.
\end{equation}
Substituting (57) and (58) into (54) yields
\begin{equation}\label{equa59}
d_{s}^{{\textrm {IC}}} = 1,
\end{equation}
which shows that a secrecy diversity gain of one is achieved by the IC-NOSS scheme. This means that given a sufficiently high SNR of $\gamma_M$, an arbitrarily small overall secrecy outage probability can be obtained by the IC-NOSS scheme, showing its advantage over the OSS and IL-NOSS methods.}}

\section{Numerical Results And Discussions}
In this section, we present numerical secrecy outage comparisons among the OSS, IL-NOSS and IC-NOSS schemes. {{In our numerical evaluation, we consider the transmission distances of ${d_{Mm}} = {d_{Ms}} = {d_{Sm}}= {d_{Me}} = {d_{Se}} = 300{{m}}$, unless otherwise stated. Since a small cell usually has a much narrower coverage than a macro cell, a distance of ${d_{Ss}} = 30{m}$ is used for the SBS-SU transmission. Also, the path loss factors of $\alpha_{Mm} = \alpha_{Ss} = \alpha_{Me} = \alpha_{Se} = 2.5$ are used, while a higher path loss factor of $\alpha_{Ms} = \alpha_{Sm} = 3$ is assumed for the cross-interference channels between the macro cell and small cell, considering that the small cell is often deployed in a shadowed area e.g. in-building places of the macro cell. Moreover, all the receivers of MU, SU and E are assumed to experience the same small-scale fading gain of $\delta _{Mm}^2 = \delta _{Ss}^2 = \delta _{Ms}^2 = \delta _{Sm}^2 = \delta _{Me}^2 = \delta _{Se}^2 = 1$. Additionally, an SNR of $\gamma_M = 100$dB, a secrecy data rata of $R_M^s =R_S^s = 1{\textrm{bit/s/Hz}}$, $\alpha  = 0.5$, and an SMR of $\beta = 0.5$ are assumed, unless otherwise mentioned.}}

\begin{figure}
\centering
\includegraphics[scale = 0.75]{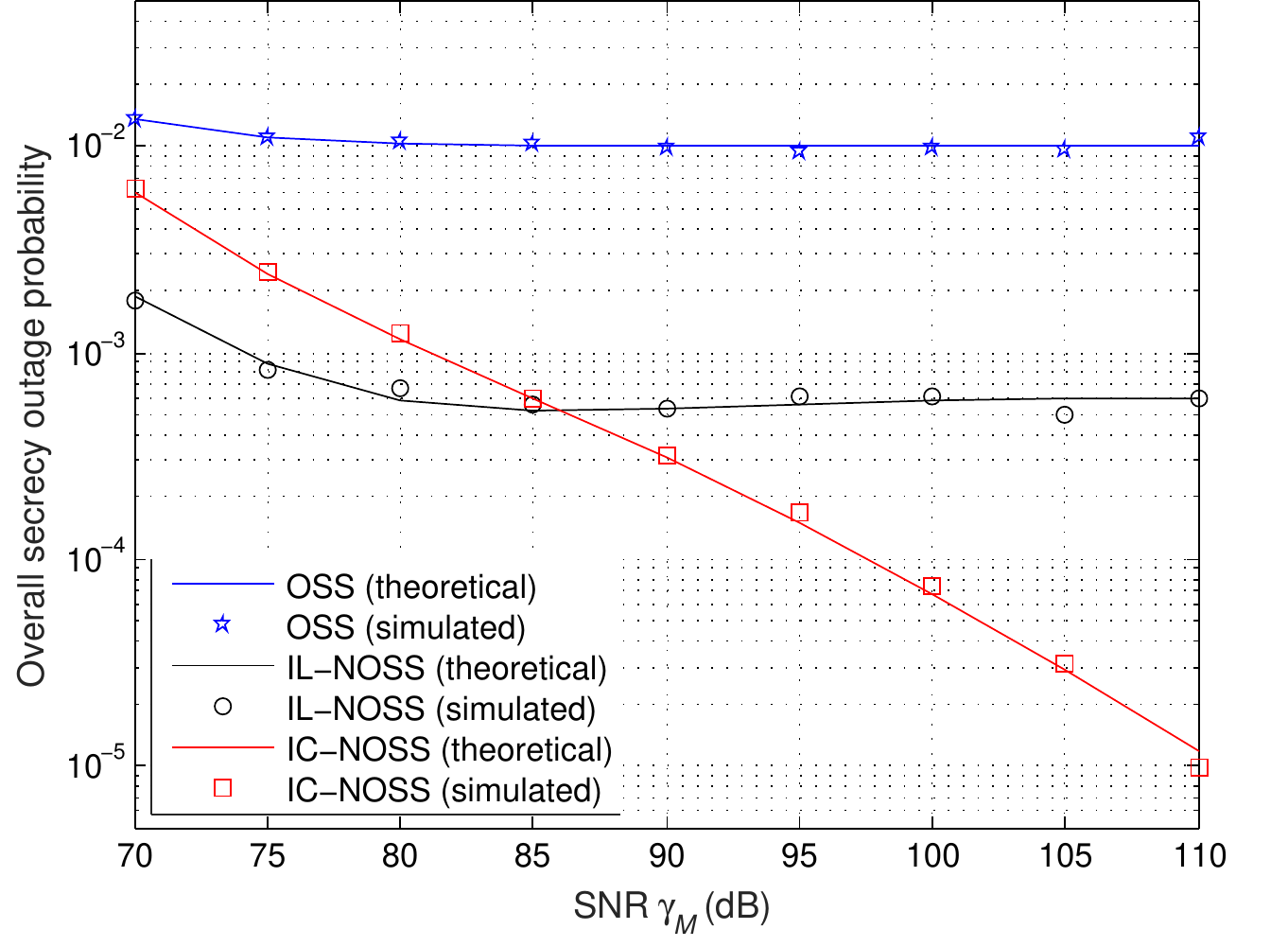}
\caption{Overall secrecy outage probability versus SNR $\gamma_M$ of the OSS, IL-NOSS and IC-NOSS schemes.}
\label{fig 2}
\end{figure}
Fig. 3 shows the overall secrecy outage probability versus SNR $\gamma_M$ for the OSS, IL-NOSS and IC-NOSS schemes, where both the theoretical and simulated secrecy outage probabilities are given. To be specific, theoretical secrecy outage curves of Fig. 3 are plotted by using (20), (28) and (40), whereas simulated secrecy outage results are obtained through Mento-Carlo simulations. As seen from Fig. 3, the theoretical and simulated secrecy outage results of OSS and IL-NOSS match well with each other.
{{It can also be observed from Fig. 3 that as the SNR increases, the secrecy outage probabilities of OSS and IL-NOSS schemes gradually converges to respective secrecy outage floors, whereas the overall secrecy outage probability of IC-NOSS scheme always decreases significantly. The occurrence of secrecy outage floors for OSS and IL-NOSS is resulted from the dominant effect of mutual interference in the high SNR region, which is, however, compensated and alleviated by using a specially-designed signal of $x_m$ as given by (2) in the IC-NOSS scheme. Furthermore, the secrecy outage performance of IC-NOSS scheme is slightly worse than that of IL-NOSS method in the low SNR region. This is because that the specially-designed signal used to cancel out the interference received at the macro cell consumes an extra transmit power, which dominates the secrecy outage degradation when the transmit power resource is very limited in the low SNR region. }}

\begin{figure}
\centering
\includegraphics[scale = 0.75]{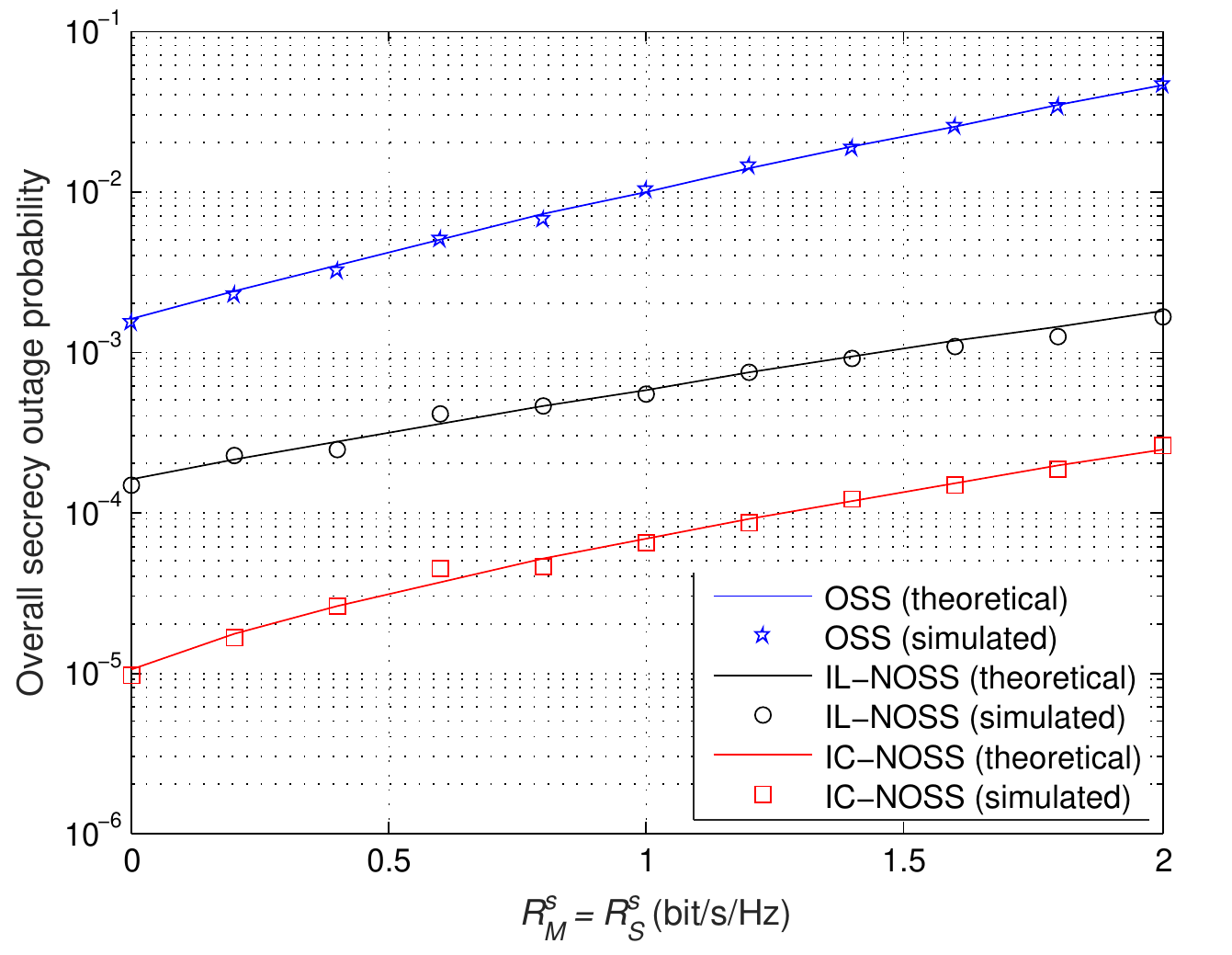}
\caption{Overall secrecy outage probability versus secrecy rate of the OSS, IL-NOSS and IC-NOSS schemes.}
\label{fig 2}
\end{figure}
Fig. 4 shows the overall secrecy outage probability versus secrecy rates of $R_M^s$ and $R_S^s$ for the OSS, IL-NOSS and IC-NOSS schemes. It can be seen from Fig. 4 that the theoretical and simulated secrecy outage probabilities of OSS, IL-NOSS and IC-NOSS schemes match well with each other, which further validates our secrecy outage probability analysis. Fig. 4 also demonstrates that as the secrecy rates of $R_M^s$ and $R_S^s$ increase, the secrecy outage probabilities of OSS, IL-NOSS and IC-NOSS schemes increase accordingly. Moreover, the overall secrecy outage probability of IC-NOSS scheme is always lower than that of the OSS and IL-NOSS methods, showing the secrecy outage benefit achieved by IC-NOSS compared with OSS and IL-NOSS.

\begin{figure}
\centering
\includegraphics[scale = 0.75]{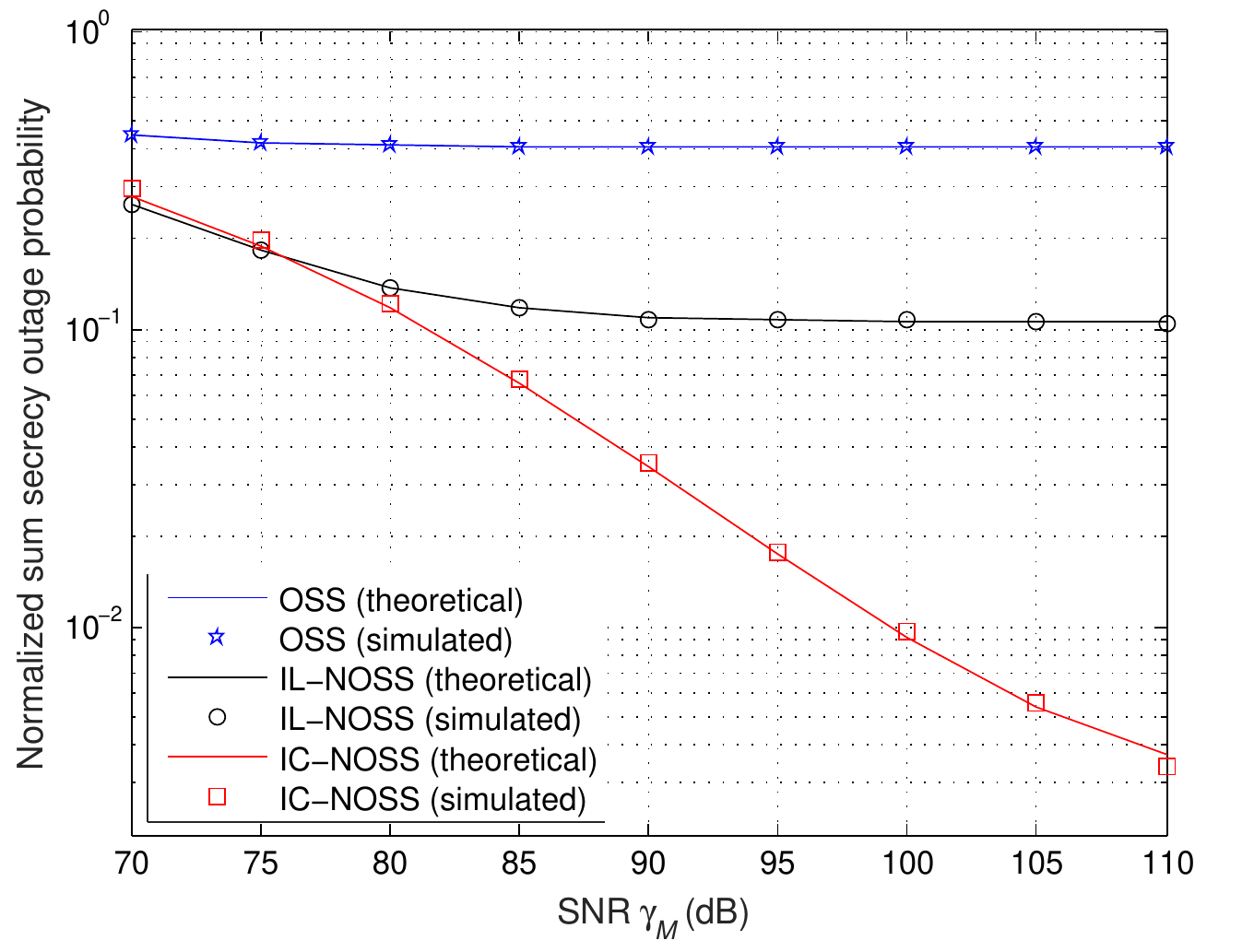}
\caption{{{Normalized sum secrecy outage probability versus SNR $\gamma_M$ of the OSS, IL-NOSS and IC-NOSS schemes, where the normalized sum secrecy outage probability is given by a mean of individual secrecy outage probabilities of the small cell and macro cell.}}}
\label{fig 2}
\end{figure}

\begin{figure}
\centering
\includegraphics[scale = 0.75]{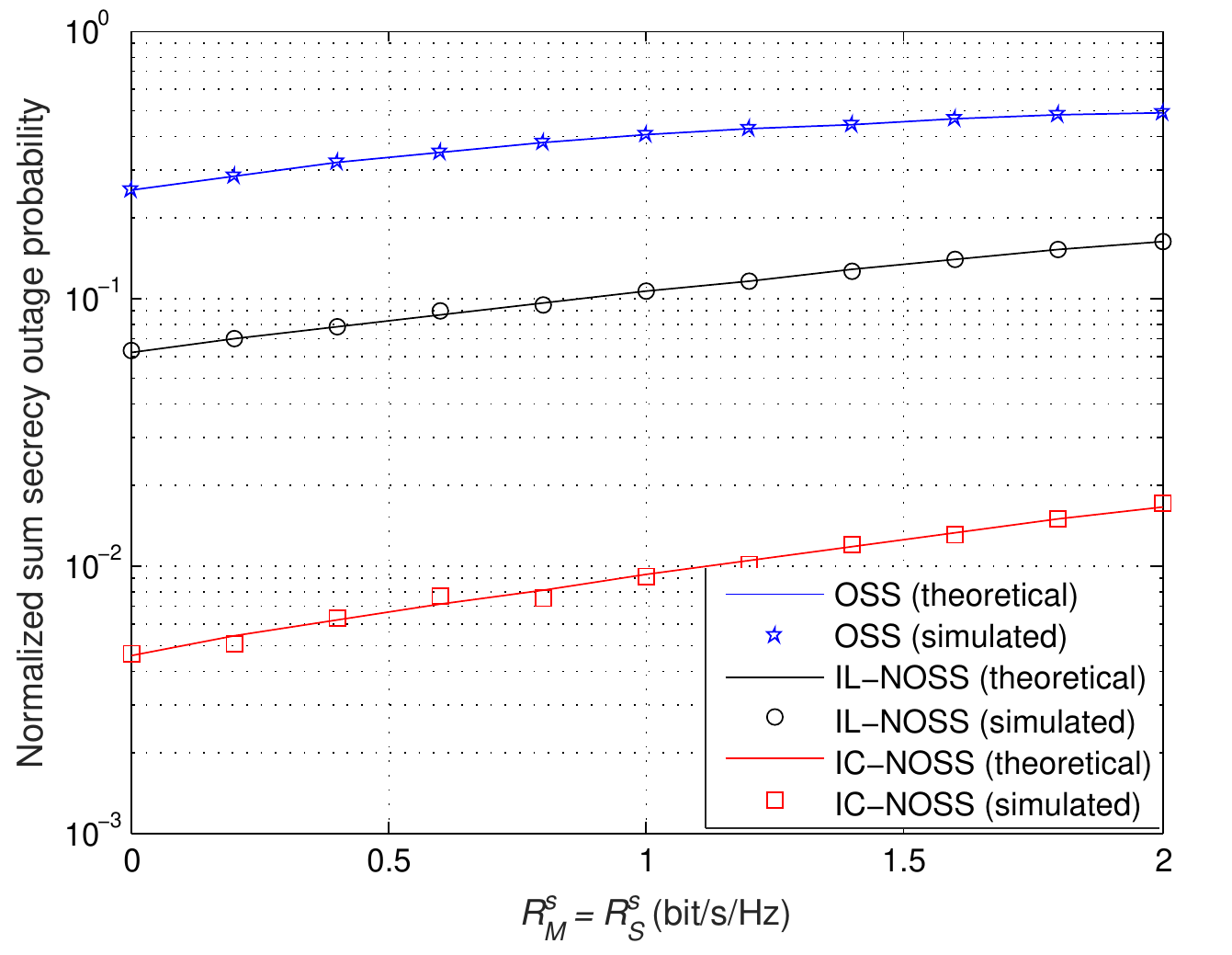}
\caption{{{Normalized sum secrecy outage probability versus secrecy rate of the OSS, IL-NOSS and IC-NOSS schemes, where the normalized sum secrecy outage probability is given by a mean of individual secrecy outage probabilities of the small cell and macro cell.}}}
\label{fig 2}
\end{figure}

{{In Fig. 5, we show the normalized sum secrecy outage probability versus SNR $\gamma_M$ of the OSS, IL-NOSS and IC-NOSS schemes, where the normalized sum secrecy outage probability is defined as a mean of individual secrecy outage probabilities of the small cell and macro cell. As shown from Fig. 5, the IC-NOSS scheme notably outperforms the OSS and IL-NOSS methods in terms of the normalized sum secrecy outage probability. Moreover, as the SNR increases, the secrecy outage advantage of IC-NOSS over OSS and IL-NOSS becomes more significant. Fig. 6 further demonstrates the normalized sum secrecy outage probability versus secrecy rate of the OSS, IL-NOSS and IC-NOSS schemes. It can be observed from Fig. 6 that the normalized sum secrecy outage probability of IC-NOSS is substantially lower than that of OSS and IL-NOSS across the whole region of secrecy rate. This also validates the effectiveness of exploiting the interference cancelation mechanism adopted in the IC-NOSS scheme against the eavesdropper.}}


\section{Conclusion}
In this paper, we studied the secrecy outage performance for a heterogeneous cellular network consisting of a macro cell and a small cell in the presence of a common eavesdropper, which attempts to tap confidential transmissions of both the macro cell and small cell. We derived closed-form expressions for an overall secrecy outage probability of OSS, IL-NOSS and IC-NOSS schemes over Rayleigh fading channels. Moreover, an asymptotic secrecy outage probability analysis was carried out for the OSS, IL-NOSS and IC-NOSS schemes to characterize their secrecy diversity gains. It was shown that the IC-NOSS scheme achieves a secrecy diversity gain of one, however the OSS and IL-NOSS methods obtain the same secrecy diversity gain of zero only. Numerical results also demonstrated that the IC-NOSS scheme generally performs better than the OSS and IL-NOSS methods in terms of the overall secrecy outage probability, especially with an increasing SNR.

{{It is pointed out that in the IC-NOSS scheme, a specially-designed signal was employed for canceling out the interference received at the macro cell as well as for confusing the eavesdropper, which, however, consumes extra transmit power resources that could be used for transmitting an information-bearing signal. An interesting extension is thus to explore the power allocation between the specially-designed signal and information-bearing signal for the sake of minimizing the secrecy outage probability. Also, this paper only considered a specific heterogeneous cellular network with a single macro cell and small cell, which can be extended to a general scenario consisting of multiple MBSs and SBSs with the aid of base station pairing and grouping. Moreover, the perfect CSI knowledge was assumed to be available in carrying out the design of interference cancelation mechanism. It is of particular interest to further examine the impact of channel estimation errors and feedback delay on the secrecy performance of our IC-NOSS scheme, which is left for future work.}}

\section*{Appendix A}
\section*{Derivation of (25)}
Denoting $U = \frac{{{{\left| {{h_{Mm}}} \right|}^2}}}{{{\gamma _S}{{\left| {{h_{Sm}}} \right|}^2} + 1}}$ and $V = \frac{{{{\left| {{h_{Me}}} \right|}^2}}}{{{\gamma _S}{{\left| {{h_{Se}}} \right|}^2} + 1}}$, we can rewrite $P_{{\textrm{out,}}M}^{{\textrm{IL}}}$ from (24) as
\begin{equation}\renewcommand\theequation{A.1}
{P_{{\textrm{out,}}M}^{{\textrm{IL}}} = \Pr [ {U < {\Delta _M} + ( {{\gamma _M}{\Delta _M} + 1} )V} ].}
\end{equation}
Noting that ${\left| {{h_{Mm}}} \right|^2}$ and ${\left| {{h_{Sm}}} \right|^2}$ are independent exponentially distributed random variables with respective means of $\sigma _{Mm}^2$ and $\sigma _{Sm}^2$, we can obtain a cumulative distribution function (CDF) of $U$ as
\begin{equation}\renewcommand\theequation{A.2}
\Pr (U <u )  =1-\frac{\sigma _{Mm}^{2}}{{{\gamma }_{S}}\sigma _{Sm}^{2}u+\sigma _{Mm}^{2}}\exp ( -\frac{u}{\sigma _{Mm}^{2}} ), \\
\end{equation}
for $u>0$. Meanwhile, since ${\left| {{h_{Me}}} \right|^2}$ and ${\left| {{h_{Se}}} \right|^2}$ are independent exponentially distributed random variables with respective means of $\sigma _{Me}^2$ and $\sigma _{Se}^2$, the CDF of $V$ can be given by
\begin{equation}\renewcommand\theequation{A.3}
\Pr (V < v ) = 1 - \frac{{\sigma _{Me}^2}}{{{\gamma _S}\sigma _{Se}^2v + \sigma _{Me}^2}}\exp ( { - \frac{v}{{\sigma _{Me}^2}}} ),
\end{equation}
from which its probability density function (PDF) is obtained as
\begin{equation}\renewcommand\theequation{A.4}
  p_V( v )=[ \frac{{{\gamma }_{S}}\sigma _{Me}^{2}\sigma _{Se}^{2}}{{{( {{\gamma }_{S}}\sigma _{Se}^{2}v+\sigma _{Me}^{2} )}^{2}}}+\frac{1}{{{\gamma }_{S}}\sigma _{Se}^{2}v+\sigma _{Me}^{2}} ] \exp ( -\frac{v}{\sigma _{Me}^{2}} ).
\end{equation}

Substituting (A.2) and (A.4) into (A.1) yields
\begin{equation}\renewcommand\theequation{A.5}
{\begin{split}
&P_{\textrm{out,}M}^{\textrm{IL}}=\int_{0}^{\infty }{\left( 1-\frac{\sigma _{Mm}^{2}}{[ {{\Delta }_{M}}+( {{\gamma }_{M}}{{\Delta }_{M}}+1 )v ]{{\gamma }_{S}}\sigma _{Sm}^{2}+\sigma _{Mm}^{2}}\exp [ -\frac{{{\Delta }_{M}}+( {{\gamma }_{M}}{{\Delta }_{M}}+1 )v}{\sigma _{Mm}^{2}} ] \right)} \\
& \quad\quad\times \left( \frac{{{\gamma }_{S}}\sigma _{Me}^{2}\sigma _{Se}^{2}}{{{( v{{\gamma }_{S}}\sigma _{Se}^{2}+\sigma _{Me}^{2} )}^{2}}}+\frac{1}{v{{\gamma }_{S}}\sigma _{Se}^{2}+\sigma _{Me}^{2}} \right)\exp ( -\frac{v}{\sigma _{Me}^{2}} )dv \\
 & =\int_{0}^{\infty }{\left( 1-\frac{1}{{{b}_{1}}( {{e}_{1}}+{{f}_{1}}v )+1}\exp [ -{{a}_{1}}( {{e}_{1}}+{{f}_{1}}v ) ] \right)}\left( \frac{{{d}_{1}}}{{{( {{d}_{1}}v+1 )}^{2}}}+\frac{{{c}_{1}}}{{{d}_{1}}v+1} \right)\exp ( -{{c}_{1}}v )dv, \\
\end{split}}
\end{equation}
where the used parameters are specified in Table I. Using (3.352) and (3.353) of [44], we can obtain $P_{{\textrm{out,}}M}^{{\textrm{IL}}}$ from (A.5) as
\begin{equation}\renewcommand\theequation{A.6}
{P_{{\textrm{out,}}M}^{{\textrm{IL}}}=1-\exp ( -{{a}_{1}}{{e}_{1}} )[-g_1\exp(h_1)Ei(-h_1)+k_1\exp(i_1)Ei(-i_1)+j_1 ]},
\end{equation}
for the case of $b_1f_1 \ne (b_1e_1+1){d_1}$. Besides, using (3.353) of [44], we arrive at
\begin{equation}\renewcommand\theequation{A.7}
{\begin{split}
P_{{\textrm{out,}}M}^{{\textrm{IL}}} =1 - \exp ( { - {a_1}{e_1}} )[ l_1+\frac{d_1}{2b_1f_1}+l_1i_1\exp(i_1)Ei(-i_1)],
\end{split}}
\end{equation}
for $b_1f_1 =(b_1e_1+1){d_1}$. Finally, combining (A.6) and (A.7) yields (13).

\section*{Appendix B}
\section*{Derivation of (34)}
{{Denoting $X = \frac{{|{h_{Mm}}{|^2}}}{{\sigma _{Mm}^2}}$, $Y = {\gamma _M}|{h_{Sm}}{|^2}$, $Z = \frac{{|{h_{Se}} {|^2}}}{{|{h_{Me}}{|^2}}}$, we can rewrite (33) as
\begin{equation}\renewcommand\theequation{B.1}
P_{{\textrm{out,}}M}^{{\textrm{IC}}} = \Pr \left( {Y + \sigma _{Mm}^2{\gamma _M}XZ < \frac{{{2^{R_M^s}}}}{{\beta X}}} \right).
\end{equation}
Since ${|{h_{Me}} {|^2}}$ and ${|{h_{Se}} {|^2}}$ are independent exponentially distributed random variables with respective means of $\sigma^2_{Me}$ and $\sigma^2_{Se}$, a CDF of $Z$ is obtained as
\begin{equation}\renewcommand\theequation{B.2}
\Pr \left( {Z < z} \right) = \Pr (\frac{{|{h_{Se}}{|^2}}}{{|{h_{Me}}{|^2}}} < z) = \frac{{\sigma _{Me}^2z}}{{\sigma _{Se}^2 + \sigma _{Me}^2z}},
\end{equation}
from which its PDF is given by
\begin{equation}\renewcommand\theequation{B.3}
p_Z(z) = \frac{{\sigma _{Me}^2\sigma _{Se}^2}}{{{{(\sigma _{Se}^2 + \sigma _{Me}^2z)}^2}}}.
\end{equation}
Using (B.3) and noting ${|{h_{Mm}} {|^2}}$ and ${|{h_{Sm}} {|^2}}$ are independent exponentially distributed random variables with respective means of $\sigma^2_{Mm}$ and $\sigma^2_{Sm}$, we have
\begin{equation}\renewcommand\theequation{B.4}
P_{{\textrm{out,}}M}^{{\textrm{IC}}} = \int_0^\infty  {f(x)\exp ( - x)dx} ,
\end{equation}
where $f(x)$ is given by
\begin{equation}\renewcommand\theequation{B.5}
\begin{split}
f(x)  = \int_0^{\frac{{{2^{R_M^s}}}}{{\sigma _{Mm}^2\beta {\gamma _M}{x^2}}}} {\frac{{\sigma _{Me}^2\sigma _{Se}^2}}{{{{(\sigma _{Se}^2 + \sigma _{Me}^2z)}^2}}}[1 - \exp ( - \frac{{{2^{R_M^s}}}}{{\sigma _{Sm}^2\beta {\gamma _M}x}} + \frac{{\sigma _{Mm}^2xz}}{{\sigma _{Sm}^2}})]dz} .
\end{split}
\end{equation}
Letting $z =  - \frac{{\sigma _{Sm}^2}}{{\sigma _{Mm}^2x}}t - \frac{{\sigma _{Se}^2}}{{\sigma _{Me}^2}}$, we can rewrite (B.5) as
\begin{equation}\renewcommand\theequation{B.6}
\begin{split}
f(x) = \frac{{{2^{R_M^s}}}}{{\sigma _{Sm}^2\beta {\gamma _M}{\varphi _x}x}} - {\phi _x}\exp ( - {\varphi _x})\int_{ - {\varphi _x}}^{ - {\phi _x}} {\frac{1}{{{t^2}}}\exp ( - t)dt} ,
\end{split}
\end{equation}
wherein ${\varphi_x} = \frac{{\sigma _{Mm}^2\sigma _{Se}^2\beta {\gamma _M}{x^2} + \sigma _{Me}^2{2^{R_M^s}}}}{{\sigma _{Sm}^2\sigma _{Me}^2\beta {\gamma _M}x}}$ and ${\phi_x}  = \frac{{\sigma _{Mm}^2\sigma _{Se}^2x}}{{\sigma _{Sm}^2\sigma _{Me}^2}}$.
Combining (B.4) and (B.6) gives (34).}}

\section*{Appendix C}
\section*{Derivation of (55) and (56)}
{{Using (33) and letting $X' = |{h_{Mm}}{|^2}$, $Y' = |{h_{Sm}}{|^2}$, and $Z = {{|{h_{Se}} {|^2}}}/{{|{h_{Me}}{|^2}}}$, we have
\begin{equation}\renewcommand\theequation{C.1}
P_{{\textrm{out,}}M}^{{\textrm{IC}}} = \Pr \left[ {X'(Y' + X'Z) < \frac{{\sigma _{Mm}^2{2^{R_M^s}}}}{{\beta {\gamma _M}}}} \right],
\end{equation}
where $\beta=\gamma_S/\gamma_M$. It is pointed out that $X'$ and $Y'$ are independent exponentially distributed random variables with respective means of $\sigma^2_{Mm}$ and $\sigma^2_{Sm}$. Moreover, the PDF of $Z$ denoted by $p_Z(z)$ is given by (B.3). Considering an inequality of $Y' + X'Z \le 2\max (Y',X'Z)$, we obtain a lower bound on $P_{{\textrm{out,}}M}^{{\textrm{IC}}}$ as
\begin{equation}\renewcommand\theequation{C.2}
P_{{\textrm{out,}}M}^{{\textrm{IC}}} \ge \underbrace {\Pr \left[ {2X'\max (Y',X'Z) < \frac{{\sigma _{Mm}^2{2^{R_M^s}}}}{{\beta {\gamma _M}}}} \right]}_{P_{{\textrm{out,}}M}^{{\rm{IC,low}}}},
\end{equation}
from which the lower bound ${P_{{\textrm{out,}}M}^{{\rm{IC,low}}}}$ is given by
\begin{equation}\renewcommand\theequation{C.3}
P_{{\textrm{out,}}M}^{{\rm{IC,low}}} = \underbrace {\Pr \left( {{X'^2}Z < \frac{{{\delta _M}}}{{{\gamma _M}}},Y' < X'Z} \right)}_{{P_1}} + \underbrace {\Pr \left( {X'Y' < \frac{{{\delta _M}}}{{{\gamma _M}}},Y' > X'Z} \right)}_{{P_2}},
\end{equation}
where ${\delta _M} = {{\sigma _{Mm}^2{2^{R_M^s}}}}/({{2\beta }})$. From (C.3), we have
\begin{equation}\renewcommand\theequation{C.4}
{P_1} = \Pr \left( {\frac{Y'}{Z} < X',X' < \sqrt {\frac{{{\delta _M}}}{{{\gamma _M}Z}}} } \right) = \int_0^\infty  {{p_Z}(z){f_1}(z)dz},
\end{equation}
where $p_Z(z)$ is the PDF of $Z$ as shown in (B.3) and ${{f_1}(z)}$ is given by
\begin{equation}\renewcommand\theequation{C.5}
\begin{split}
{f_1}(z) &= \int_0^{\sqrt {\frac{{{\delta _M}z}}{{{\gamma _M}}}} } {\frac{1}{{\sigma _{Sm}^2}}\exp ( - \frac{y}{{\sigma _{Sm}^2}})dy\int_{\frac{y}{z}}^{\sqrt {\frac{{{\delta _M}}}{{{\gamma _M}z}}} } {\frac{1}{{\sigma _{Mm}^2}}\exp ( - \frac{x}{{\sigma _{Mm}^2}})dx} }\\
& = \frac{{\sigma _{Mm}^2z}}{{\sigma _{Mm}^2z + \sigma _{Sm}^2}}[1 - \exp ( - {\lambda _M} - {\lambda _S})] - \exp ( - {\lambda _M})[1 - \exp ( - {\lambda _S})],
\end{split}
\end{equation}
where ${\lambda _M} = \frac{1}{{\sigma _{Mm}^2}}\sqrt {\frac{{{\delta _M}}}{{{\gamma _M}z}}}
$ and ${\lambda _S} = \frac{1}{{\sigma _{Sm}^2}}\sqrt {\frac{{{\delta _M}z}}{{{\gamma _M}}}}$. It can be observed that both ${\lambda _M} $ and ${\lambda _S} $ approach to zero for $\gamma_M \to \infty$. Using the Taylor series expansion and ignoring the high order infinitesimal, we arrive at
\begin{equation}\renewcommand\theequation{C.6}
\begin{split}
1 - \exp ( - {\lambda _M} - {\lambda _S}) &= {\lambda _M} + {\lambda _S} - \frac{1}{2}{({\lambda _M} + {\lambda _S})^2}\\
&= \frac{{\sigma _{Mm}^2z + \sigma _{Sm}^2}}{{\sigma _{Mm}^2z}}{\lambda _S} - \frac{1}{2}{(\frac{{\sigma _{Mm}^2z + \sigma _{Sm}^2}}{{\sigma _{Mm}^2z}})^2}\lambda _S^2,
\end{split}
\end{equation}
where the second equation is obtained by using $ {\lambda _M} + {\lambda _S} = \frac{{\sigma _{Mm}^2z + \sigma _{Sm}^2}}{{\sigma _{Mm}^2z}}{\lambda _S}$. Similarly, we have
\begin{equation}\renewcommand\theequation{C.7}
\exp ( - {\lambda _M}) = 1 - {\lambda _M},
\end{equation}
and
\begin{equation}\renewcommand\theequation{C.8}
1 - \exp ( - {\lambda _S}) = {\lambda _S} - \frac{1}{2}\lambda _S^2,
\end{equation}
for $\gamma_M \to \infty$. Substituting (C.6)-(C.8) into (C.5) yields
\begin{equation}\renewcommand\theequation{C.9}
{f_1}(z) =  - \frac{1}{2}\frac{{\sigma _{Sm}^2}}{{\sigma _{Mm}^2z}}\lambda _S^2 + {\lambda _M}{\lambda _S} - \frac{1}{2}{\lambda _M}\lambda _S^2,
\end{equation}
for $\gamma_M \to \infty$. Since ${\lambda _M}\lambda _S^2$ is a higher-order infinitesimal than $\lambda _S^2$ and ${\lambda _M}{\lambda _S}$, we can ignore the term of (C.9) and obtain
\begin{equation}\renewcommand\theequation{C.10}
{f_1}(z) = \frac{{{\delta _M}}}{{2\sigma _{Mm}^2\sigma _{Sm}^2{\gamma _M}}},
\end{equation}
which is obtained by using ${\lambda _M} = \frac{1}{{\sigma _{Mm}^2}}\sqrt {\frac{{{\delta _M}}}{{{\gamma _M}z}}}$ and ${\lambda _S} = \frac{1}{{\sigma _{Sm}^2}}\sqrt {\frac{{{\delta _M}z}}{{{\gamma _M}}}} $. Substituting ${\delta _M} = \frac{{\sigma _{Mm}^2{2^{R_M^s}}}}{{2\beta }}$ and (C.10) into (C.4) gives
\begin{equation}\renewcommand\theequation{C.11}
{P_1} = \frac{{{2^{R_M^s}}}}{{4\sigma _{Sm}^2\beta {\gamma _M}}},
\end{equation}
for $\gamma_M \to \infty$. Moreover, from (C.3), $P_2$ is obtained as
\begin{equation}\renewcommand\theequation{C.12}
{P_2} = \Pr \left( {X'Z < Y',Y' < \frac{{{\delta _M}}}{{{\gamma _M}X'}}} \right) = \int_0^\infty  {{p_Z}(z){f_2}(z)dz} ,
\end{equation}
where ${f_2}(z) $ is given by
\begin{equation}\renewcommand\theequation{C.13}
\begin{split}
{f_2}(z) &= \int_0^{\sqrt {\frac{{{\delta _M}}}{{{\gamma _M}z}}} } {\frac{1}{{\sigma _{Mm}^2}}\exp ( - \frac{x}{{\sigma _{Mm}^2}})dx\int_{xz}^{\frac{{{\delta _M}}}{{{\gamma _M}x}}} {\frac{1}{{\sigma _{Sm}^2}}\exp ( - \frac{y}{{\sigma _{Sm}^2}})dy} } \\
& = \int_0^1 {{\lambda _M}[\exp ( - {\lambda _M}t - {\lambda _S}t) - \exp ( - {\lambda _M}t - \frac{{{\lambda _S}}}{t})]dt}  \\
\end{split}
\end{equation}
wherein ${\lambda _M} = \frac{1}{{\sigma _{Mm}^2}}\sqrt {\frac{{{\delta _M}}}{{{\gamma _M}z}}}$. Denoting ${\lambda _S} = \frac{1}{{\sigma _{Sm}^2}}\sqrt {\frac{{{\delta _M}z}}{{{\gamma _M}}}} $ and substituting $x = \sigma _{Mm}^2{\lambda _M}t$, we can rewrite (C.13) as
\begin{equation}\renewcommand\theequation{C.14}
\begin{split}
{f_2}(z)  = \underbrace {\int_0^1 {{\lambda _M}\exp ( - {\lambda _M}t - {\lambda _S}t)dt} }_{{f_{2,1}}(z)} - \underbrace {\int_0^1 {{\lambda _M}\exp ( - {\lambda _M}t - \frac{{{\lambda _S}}}{t})dt} }_{{f_{2,2}}(z)},  \\
\end{split}
\end{equation}
from which ${f_{2,1}}(z) $ is given by
\begin{equation}\renewcommand\theequation{C.15}
{f_{2,1}}(z) = \frac{{{\lambda _M}}}{{{\lambda _M} + {\lambda _S}}}[1 - \exp ( - {\lambda _S} - {\lambda _M})].
\end{equation}
Combining (C.6) and (C.15) and ignoring the high-order infinitesimal yield
\begin{equation}\renewcommand\theequation{C.16}
{f_{2,1}}(z) = {\lambda _M},
\end{equation}
for $\gamma_M \to \infty$. Besides, using $t = \frac{{{\lambda _S}}}{x}$, we can obtain ${f_{2,2}}(z) $ from (C.14) as
\begin{equation}\renewcommand\theequation{C.17}
\begin{split}
 {f_{2,2}}(z)& = \int_{{\lambda _S}}^\infty  {\frac{{{\lambda _M}{\lambda _S}}}{{{x^2}}}\exp ( - \frac{{{\lambda _M}{\lambda _S}}}{x} - x)dx}  \\
&= \int_{{\lambda _S}}^\infty  {\frac{{{\lambda _M}{\lambda _S}}}{{{x^2}}}\exp ( - x)dx}  \\
&= {\lambda _M}{\lambda _S}[\exp ( - {\lambda _S})\frac{1}{{{\lambda _S}}} - Ei({\lambda _S})], \\
 \end{split}
\end{equation}
where the second equation is obtained by ignoring the high order infinitesimal ${\frac{{{\lambda _M}{\lambda _S}}}{x}}$ for $\gamma_M \to \infty$. Substituting $\exp ( - {\lambda _S}) = 1 - {\lambda _S}$ into (C.17) and ignoring high order infinitesimals yield
\begin{equation}\renewcommand\theequation{C.18}
{f_{2,2}}(z) = {\lambda _M} - {\lambda _M}{\lambda _S}Ei({\lambda _S}).
\end{equation}
Substituting (C.16) and (C.18) into (C.14) gives
\begin{equation}\renewcommand\theequation{C.19}
{f_2}(z) = \frac{{{\delta _M}}}{{\sigma _{Sm}^2\sigma _{Mm}^2{\gamma _M}}}Ei(\frac{1}{{\sigma _{Sm}^2}}\sqrt {\frac{{{\delta _M}z}}{{{\gamma _M}}}} ),
\end{equation}
where ${\lambda _M} = \frac{1}{{\sigma _{Mm}^2}}\sqrt {\frac{{{\delta _M}}}{{{\gamma _M}z}}}
$ and ${\lambda _S} = \frac{1}{{\sigma _{Sm}^2}}\sqrt {\frac{{{\delta _M}z}}{{{\gamma _M}}}}$ are used. Following (5.1.20) of [45], we have
\begin{equation}\renewcommand\theequation{C.20}
\frac{1}{2}\exp ( - x)\ln (1 + \frac{2}{x}) \le Ei(x) \le \exp ( - x)\ln (1 + \frac{1}{x}),
\end{equation}
for $x>0$. Combining (C.19) and (C.20) yields
\begin{equation}\renewcommand\theequation{C.21}
\dfrac{{{\delta _M}\ln ({\gamma _M})}}{{4\sigma _{Sm}^2\sigma _{Mm}^2{\gamma _M}}} \le \mathop {\lim }\limits_{{\gamma _M} \to \infty } {f_2}(z) \le \dfrac{{{\delta _M}\ln ({\gamma _M})}}{{2\sigma _{Sm}^2\sigma _{Mm}^2{\gamma _M}}}.
\end{equation}
Substituting ${\delta _M} = \frac{{\sigma _{Mm}^2{2^{R_M^s}}}}{{2\beta }}$ and (C.21) into (C.12) gives
\begin{equation}\renewcommand\theequation{C.22}
\frac{{{2^{R_M^s}}\ln ({\gamma _M})}}{{8\sigma _{Sm}^2\beta {\gamma _M}}} \le \mathop {\lim }\limits_{{\gamma _M} \to \infty } {P_2} \le \frac{{{2^{R_M^s}}\ln ({\gamma _M})}}{{4\sigma _{Sm}^2\beta {\gamma _M}}}.
\end{equation}
Combining (C.11) and (C.22) with (C.2) and (C.3) as well as ignoring high order infinitesimals, we arrive at
\begin{equation}\renewcommand\theequation{C.23}
P_{{\textrm{out,}}M}^{{\textrm{IC}}} \ge \frac{{{2^{R_M^s}}\ln ({\gamma _M})}}{{8\sigma _{Sm}^2\beta {\gamma _M}}},
\end{equation}
which gives a lower bound of $P_{{\textrm{out,}}M}^{{\textrm{IC}}}$.}}

{{In addition, using an inequality of $Y' + X'Z \ge \max (Y',X'Z)$, we obtain an upper bound on $P_{{\textrm{out,}}M}^{{\textrm{IC}}}$ from (C.1) as
\begin{equation}\renewcommand\theequation{C.24}
P_{{\textrm{out,}}M}^{{\textrm{IC}}} \le \Pr \left[ {X'\max (Y',X'Z) < \frac{{\sigma _{Mm}^2{2^{R_M^s}}}}{{\beta {\gamma _M}}}} \right].
\end{equation}
It can be observed that only an extra coefficient of $2$ is introduced in (C.2) compared to (C.24). Thus, similar to (C.11) and (C.22), one can readily have
\begin{equation}\renewcommand\theequation{C.25}
P_{{\textrm{out,}}M}^{{\textrm{IC}}} \le \frac{{{2^{R_M^s}}\ln ({\gamma _M})}}{{2\sigma _{Sm}^2\beta {\gamma _M}}},
\end{equation}
which is an upper bound of $P_{{\textrm{out,}}M}^{{\textrm{IC}}}$.}}


\begin{thebibliography}{32}
\bibitem{IEEEhowto:1}
Y. Zou, \textquotedblleft Intelligent interference exploitation for heterogeneous cellular networks against eavesdropping," \emph{IEEE J. Sel. Areas Commun.}, vol. 36, no. 7, pp. 1453-1464, Jul. 2018.

\bibitem{IEEEhowto:2}
S. Singh, H. S. Dhillon and J. G. Andrews, ``Offloading in heterogeneous networks: Modeling, analysis, and design Insights," \emph{IEEE Trans. Wirel. Commun.}, vol. 12, no. 5, pp. 2484-2497, May 2013.

\bibitem{IEEEhowto:3}
{{Y. Zou, M. Sun, J. Zhu, and H. Guo, ``Security-reliability tradeoff for distributed antenna systems in heterogeneous cellular networks," \emph{IEEE Trans. Wirel. Commun.}, vol. 17, no. 12, pp. 8444-8456, Dec. 2018.}}

\bibitem{IEEEhowto:4}
A. Ghosh \emph{et al.}, \textquotedblleft Heterogeneous cellular networks: From theory to practice," \emph{IEEE Commun. Mag.}, vol. 50, no. 6, pp. 54-64, Jun. 2012.

\bibitem{IEEEhowto:5}
R. Madan, J. Borran, A. Sampath, N. Bhushan, A. Khandekar, and T. Ji, \textquotedblleft Cell association and interference coordination in heterogeneous LTE-A cellular networks," \emph{IEEE J. Sel. Areas Commun.}, vol. 28, no. 9, pp. 1479-1489, Dec. 2010.

\bibitem{IEEEhowto:6}
S. Singh and J. G. Andrews, \textquotedblleft Joint resource partitioning and offloading in heterogeneous cellular networks," \emph{IEEE Trans. Wirel. Commun.}, vol. 13, no. 2, pp. 888-901, Feb. 2014.

\bibitem{IEEEhowto:7}
I. Hwang, B. Song, and S. S. Soliman, \textquotedblleft A holistic view on hyper-dense heterogeneous and small cell networks," \emph{IEEE Commun. Mag.}, vol. 51, no. 6, pp. 20-27, Jun. 2013.

\bibitem{IEEEhowto:8}
D. Fooladivanda and C. Rosenberg, \textquotedblleft Joint resource allocation and user association for heterogeneous wireless cellular networks," \emph{IEEE Trans. Wirel. Commun.}, vol. 12, no. 1, pp. 248-257, Jan. 2013.

\bibitem{IEEEhowto:9}
Q. Ye, B. Rong, Y. Chen, M. Al-Shalash, C. Caramanis, and J. G. Andrews, \textquotedblleft User association for load balancing in heterogeneous cellular networks," \emph{ IEEE Trans. Wirel. Commun.}, vol. 12, no. 6, pp. 2706-2716, Jun. 2013.

\bibitem{IEEEhowto:10}
M. O. Al-Kadri, Y. Deng, A. Aijaz, and A. Nallanathan, \textquotedblleft Full-duplex small cells for next generation heterogeneous cellular networks: A case study of outage and rate coverage analysis," \emph{IEEE Access}, vol. 5, pp. 8025-8038, May 2017.

\bibitem{IEEEhowto:11}
L. Tang, W. Wang, Y. Wang, and Q. Chen, \textquotedblleft An energy-saving algorithm with joint user association, clustering, and on/off strategies in dense heterogeneous networks," \emph{IEEE Access}, vol. 5, pp. 12988-13000, Jul. 2017.

\bibitem{IEEEhowto:12}
S. Navaratnarajah, M. Dianati, and M. A. Imran, "A novel load-balancing scheme for cellular-WLAN heterogeneous systems with a cell-breathing technique," \emph{IEEE Syst. J.} no. 99, pp. 1-12, Aug. 2017.

\bibitem{IEEEhowto:13}
O. Mehanna, \textquotedblleft Sharing vs. splitting spectrum in OFDMA femtocell networks," \emph{IEEE International Conference on Acoustics, Speech and Signal Processing} , pp. 4824-4828, Vancouver, BC, 2013.

\bibitem{IEEEhowto:14}
Z. Ding, Y. Liu, J. Choi, \emph{et al.}, ``Application of non-orthogonal multiple access in LTE and 5G networks," \emph{IEEE Commun. Mag.}, vol. 55, no. 2, pp. 185-191, Feb. 2017.

\bibitem{IEEEhowto:15}
S. Islam, N. Avazov, O. Dobre, and K. Kwak, ``Power-domain non-orthogonal multiple access (NOMA) in 5G systems: Potentials and challenges," \emph{IEEE Commun. Surv. \& Tut.}, vol. 19, no. 2, pp. 721-742, Oct. 2016.

\bibitem{IEEEhowto:16}
S. Mukherjee, \textquotedblleft Distribution of downlink SINR in heterogeneous cellular networks," \emph{IEEE J. Sel. Areas Commun.}, vol. 30, no. 3, pp. 575-585, Apr. 2012.

\bibitem{IEEEhowto:17}
K. Shanmugam, N. Golrezaei, A. G. Dimakis, A. F. Molisch, and G. Caire, \textquotedblleft FemtoCaching: Wireless content delivery through distributed caching helpers," \emph{IEEE Trans. Inf. Theory}, vol. 59, no. 12, pp. 8402-8413, Dec. 2013.

\bibitem{IEEEhowto:18}
D. Cao, S. Zhou, and Z. Niu, \textquotedblleft Improving the energy efficiency of two-tier heterogeneous cellular networks through partial spectrum reuse," \emph{IEEE Trans. Wirel. Commun.}, vol. 12, no. 8, pp. 4129-4141, Aug. 2013.

\bibitem{IEEEhowto:19}
Z. H. Abbas, F. Muhammad, and L. Jiao, \textquotedblleft Analysis of load balancing and interference management in heterogeneous cellular networks," \emph{IEEE Access}, vol. 5, pp. 14690-14705, Jul. 2017.

\bibitem{IEEEhowto:20}
W. Noh and K. Jang, \textquotedblleft Downlink interference control in heterogeneous cellular networks: Macroscopic and microscopic control," \emph{IEEE Trans. Veh. Tech.}, vol. 66, no. 7, pp. 5919-5932, Jul. 2017.

\bibitem{IEEEhowto:21}
Y. Zou, J. Zhu, X. Wang, and L. Hanzo, \textquotedblleft A survey on wireless security: Technical challenges, recent advances, and future trends," \emph{Proc. IEEE}, vol. 104, no. 9, pp. 1727-1765, Sep. 2016.

\bibitem{IEEEhowto:22}
Y. Zou, J. Zhu, L. Yang, Y.-.C. Liang, and Y.-D. Yao, ``Securing physical-layer communications for cognitive radio networks," \emph{IEEE Commun. Mag.}, vol. 53, no. 9, pp. 48-54, Sept. 2015.


\bibitem{IEEEhowto:23}
{{N. Yang, L. Wang, G. Geraci, M. Elkashlan, J. Yuan, and M. D. Renzo, \textquotedblleft Safeguarding 5G wireless communication networks using physical layer security," \emph{IEEE Commun. Mag.}, vol. 53, no. 4, pp. 20-27, Apr. 2015.}}

\bibitem{IEEEhowto:24}
Y. Zou, X. Wang, and W. Shen, \textquotedblleft Optimal relay selection for physical-layer security in cooperative wireless networks," \emph{IEEE J. Sel. Areas Commun.}, vol. 31, no. 10, pp. 2099-2111, Oct. 2013.

\bibitem{IEEEhowto:25}
Y. Huang, J. Wang, C. Zhong, T. Q. Duong, and G. K. Karagiannidis, ``Secure transmission in cooperative relaying networks with multiple antennas," \emph{IEEE Trans. Wirel. Commun.}, vol. 15, no. 10, pp. 6843-6856, Oct. 2016.

\bibitem{IEEEhowto:26}
{{X. Chen, C. Zhong, C. Yuen, and H.-H. Chen, ``Multi-antenna relay aided wireless physical layer security," \emph{IEEE Commun. Mag.}, vol. 53, no. 12, pp. 40-46, Dec. 2015.}}

\bibitem{IEEEhowto:27}
{{L. J. Rodriguez, N. H. Tran, T. Q. Duong, \emph{et al.}, ``Physical layer security in wireless cooperative relay networks: State of the art and beyond," \emph{IEEE Commun. Mag.}, vol. 53, no. 12, pp. 32-39, Dec. 2015.}}

\bibitem{IEEEhowto:28}
N. Yang, S. Yan, J. Yuan, R. Malaney, R. Subramanian, and I. Land, ``Artificial noise: Transmission optimization in multi-input single-output wiretap channels," \emph{IEEE Trans. Commun.,} vol. 63, no. 5, pp. 1771-1783, May 2015.


\bibitem{IEEEhowto:29}
{{D. Kapetanovic, G. Zheng, and F. Rusek, ``Physical layer security for massive MIMO: An overview on passive eavesdropping and active attacks," \emph{IEEE Commun. Mag.}, vol. 53, no. 6, pp. 21-27, Jun. 2015.}}

\bibitem{IEEEhowto:30}
{{N. Zhao, F. R. Yu, M. Li, \emph{et al.}, ``Physical layer security issues in interference alignment (IA)-based wireless networks," \emph{IEEE Commun. Mag.}, vol. 54, no. 8, pp. 162-168, Aug. 2016.}}

\bibitem{IEEEhowto:31}
H.-M. Wang, Q. Yin, and X.-G. Xia, ``Distributed beamforming for physical-layer security of two-way relay networks," \emph{IEEE Trans. Signal Process.}, vol. 60, no. 7, pp. 3532-3545, Jul. 2012.

\bibitem{IEEEhowto:32}
F. Zhu and M. Yao, \textquotedblleft Improving physical-layer security for CRNs using SINR-based cooperative beamforming," \emph{IEEE Trans. Veh. Tech.}, vol. 65, no. 3, pp. 1835-1841, Mar. 2016.

\bibitem{IEEEhowto:33}
{{C. Wang and H.-M. Wang, \textquotedblleft Physical layer security in millimeter wave cellular networks," \emph{IEEE Trans. Wirel. Commun.}, vol. 15, no. 8, pp.  5569-5585, Aug. 2016.}}

\bibitem{IEEEhowto:34}
{{Y. Zhu, L. Wang, K.-K. Wong, and R. W. Heath , \textquotedblleft Secure communications in millimeter wave ad hoc networks," \emph{IEEE Trans. Veh. Tech.}, vol. 16, no. 5, pp. 3205-3217, May. 2017.}}

\bibitem{IEEEhowto:35}
{{Y. Liu, Z. Qin, M. Elkashlan, Y. Guo, and L. Hanzo, ``Enhancing the physical layer security of non-orthogonal multiple access in large-scale networks," \emph{IEEE Trans. Wirel. Commun.}, vol. 16, no. 3, pp. 1656-1672, Mar. 2016.}}

\bibitem{IEEEhowto:36}
{{Y. Deng, L. Wang, S. Zaidi, J, Yuan, and M. Elkashlan, ``Artificial-noise aided secure transmission in large scale spectrum sharing networks," \emph{IEEE Trans. Commun.}, vol. 64, no. 5, pp. 2116-2129, May 2016.}}

\bibitem{IEEEhowto:37}
{{H.-M. Wang, T. X. Zheng, J. Yuan, D. Towsley, and M. H. Lee, \textquotedblleft Physical layer security in heterogeneous cellular networks," \emph{IEEE Trans. Commun.}, vol. 64, no. 3, pp. 1204-1219, Mar. 2016.}}

\bibitem{IEEEhowto:38}
Y. Zou, ``Physical-layer security for spectrum sharing systems," \emph{IEEE Trans. Wirel. Commun.}, vol. 16, no. 2, pp. 1319-1329, Feb. 2017.

\bibitem{IEEEhowto:39}
{{L. Wang, K,-K, Wong, S, Jin, G, Zheng, and R. W. Heath, ``A new look at physical layer security, caching, and wireless energy harvesting for heterogeneous ultra-dense networks," \emph{IEEE Commun. Mag.}, vol. 56, no. 6, pp. 49-55, Jun. 2018.}}

\bibitem{IEEEhowto:40}
{{H. Lei, C. Gao, I, Ansari, et al., ``Secrecy outage performance of transmit antenna selection for MIMO underlay cognitive radio systems over Nakagami-m channels," \emph{IEEE Trans. Veh. Tech.}, vol. 66, no. 3, pp. 2237-2250, Mar. 2017.}}

\bibitem{IEEEhowto:41}
{{S. Yan and R. Malaney, ``Location-based beamforming and physical layer security in Rician wiretap channels," \emph{IEEE Trans. Wirel. Commun.}, vol. 15, no. 4, pp. 2780-2791, Apr. 2016.}}

\bibitem{IEEEhowto:42}
{{H. Lei, I. Ansari, G. Pan, B. Alomair, M.-S. Alouini, ``Secrecy capacity analysis over $\alpha$-$\mu$ fading channels," \emph{IEEE Commun. Lett.}, vol. 21, no. 6, pp. 1445-1448, Jun. 2017.}}

\bibitem{IEEEhowto:43}
{{A. F. Molisch, \emph{Wireless Communications}, Wiley, New Jersey, USA, 2011.}}

\bibitem{IEEEhowto:44}
I. S. Gradshteyn and I. M. Ryzhik, \emph{Tables of Integrals, Series, and Products}, NY, USA: Academic Press, 2007.

\bibitem{IEEEhowto:45}
M. Abramowitz and I. Stegun, Handbook of Mathematical Functions, New York: Dover, 1964.

\end{thebibliography}
\end{document}